\documentclass{article}
\usepackage{amsfonts,amsmath,axodraw}
\begin{document}

\newcommand{\AdS}{\mathrm{AdS}}
\newcommand{\Supertwistor}{\Cset \mathrm{P}^{3|4}}
\newcommand{\dbar}{\bar{\partial}}
\newcommand{\Dbar}{\overline{D}}
\newcommand{\Tr}{\mathrm{Tr}}
\newcommand{\tlambda}{\widetilde{\lambda}}
\newcommand{\tchi}{\widetilde{\chi}}
\newcommand{\tphi}{\widetilde{\phi}}
\newcommand{\Dirac}{D\!\!\!\!\slash}

\newcommand{\Zset}{{\mathbb Z}}
\newcommand{\Cset}{{\,\,{{{^{_{\pmb{\mid}}}}\kern-.47em{\mathrm C}}}}}

\newcommand{\p}{\partial}
\newcommand{\vt}{\vartheta}
\newcommand{\vf}{\varphi}
\newcommand{\half}{\frac{1}{2}}
\newcommand{\diff}{\mathrm{d}}

\newcommand{\ra}{\rightarrow}
\newcommand{\gra}{\alpha}
\newcommand{\grl}{\lambda}
\newcommand{\gre}{\epsilon}
\newcommand{\zb}{{\bar{z}}}
\newcommand{\mn}{{\mu\nu}}
\newcommand{\Acal}{{\mathcal A}}
\newcommand{\Rcal}{{\mathcal R}}
\newcommand{\Dcal}{{\mathcal D}}
\newcommand{\Mcal}{{\mathcal M}}
\newcommand{\Ncal}{{\mathcal N}}
\newcommand{\Lcal}{{\mathcal L}}
\newcommand{\Scal}{{\mathcal S}}
\newcommand{\Wcal}{{\mathcal W}}
\newcommand{\Bcal}{\mathcal{B}}
\newcommand{\Ccal}{\mathcal{C}}
\newcommand{\Vcal}{\mathcal{V}}
\newcommand{\Ocal}{\mathcal{O}}
\newcommand{\Zb}{\overline{Z}}
\newcommand{\Urm}{{\mathrm U}}
\newcommand{\Srm}{{\mathrm S}}
\newcommand{\SO}{\mathrm{SO}}
\newcommand{\Sp}{\mathrm{Sp}}
\newcommand{\SU}{\mathrm{SU}}

\newcommand{\ip}[1]{\langle{#1}\rangle}
\newcommand{\hb}{\overline{h}}

\newcommand{\Wick}[1]{\;\,\rule[10pt]{.5pt}{2pt}\!\hspace{1.2pt}\rule[12pt]{#1}{.5pt}
\!\hspace{1.1pt}\rule[10pt]{.5pt}{2pt}\hspace{-#1}\hspace{-1.4mm}}
\newcommand{\Wicktall}[1]{\;\,\rule[10pt]{.5pt}{4pt}\!\hspace{1.2pt}\rule[14pt]{#1}{.5pt}
\!\hspace{1.1pt}\rule[10pt]{.5pt}{4pt}\hspace{-#1}\hspace{-1.4mm}}
\newcommand{\Wickunder}[1]{\;\,\rule[-6pt]{.5pt}{2pt}\!\hspace{1.2pt}\rule[-6pt]{#1}{.5pt}
\!\hspace{1.1pt}\rule[-6pt]{.5pt}{2pt}\hspace{-#1}\hspace{-1.4mm}}

\long\def\symbolfootnote[#1]#2{\begingroup%
\def\thefootnote{\fnsymbol{footnote}}\footnote[#1]{#2}\endgroup}

\numberwithin{equation}{section}

\begin{titlepage}

\begin{flushright}
 {\tt YITP-SB-04-54}
\end{flushright}

\mbox{}
\bigskip
\bigskip
\bigskip

\begin{center}
{\Large \bf Marginal Deformations of $\Ncal=4$ SYM \\ \vspace{5pt} from Open/Closed Twistor Strings}\\
\end{center}

\bigskip  
\bigskip  
\bigskip  
\bigskip 
  
\centerline{\bf Manuela Kulaxizi and Konstantinos Zoubos\symbolfootnote[1]{Present Address: Department of Physics, 
Queen Mary, University of London (k.zoubos@qmul.ac.uk)}}
 
\bigskip  
\bigskip

\centerline{\it C. N. Yang Institute for Theoretical Physics}
\centerline{\it State University of New York at Stony Brook}
\centerline{\it Stony Brook, New York 11794-3840}
\centerline{\it U. S. A.}
\bigskip

\begin{center} {\small \tt  
kulaxizi@insti.physics.sunysb.edu \\ kzoubos@insti.physics.sunysb.edu}  
\end{center}
  
\bigskip  
\bigskip  
\bigskip  

\begin{center} {\bf Abstract}\end{center}

\vskip 4pt  We investigate how the marginal deformations of $\Ncal=4$ supersymmetric Yang--Mills 
theory (analysed in particular by Leigh and Strassler) arise within B--model topological string 
theory on supertwistor space $\Supertwistor$. This is achieved by turning on a certain closed 
string background in the fermionic directions. Through a specific open/closed correlation function, 
this mode induces a correction to holomorphic Chern--Simons theory, corresponding to the self--dual 
part of the chiral operator added on the gauge theory side.
The effect of the deformation is interpreted as non--anticommutativity between some of the 
odd coordinates of $\Supertwistor$. Motivated by this, we extend the twistor formalism for 
calculating MHV amplitudes in $\Ncal=4$ SYM to these $\Ncal=1$ theories by introducing a suitable 
star product between the wavefunctions. We check that our prescription yields the expected results 
to linear order in the deformation parameter.

\end{titlepage}

\setcounter{footnote}{0}  
\noindent

\tableofcontents

\section{Introduction}

 The $\Ncal=4$ supersymmetric Yang--Mills (SYM) theory in four dimensions has many remarkable properties, 
which have made it a very useful model in explorations of links between gauge theory and 
string theory in recent years. Chief among these properties are the exact quantum conformal 
invariance of the theory and its $\mathrm{SL}(2,\Zset)$ duality symmetry (an extension of Montonen--Olive
duality). 

 It is even more remarkable that these properties do not crucially depend of the high amount of
supersymmetry, but are in fact shared by a large class of gauge theories
with $\Ncal=1$ supersymmetry, of which the $\Ncal=4$ theory is a special case. In particular, as 
first shown systematically by Leigh and Strassler \cite{LeighStrassler95}, there exists a two--complex
dimensional moduli space of exactly marginal deformations of the $\Ncal=4$ theory. Each point on this
moduli space corresponds to a \emph{finite}, and thus conformally invariant, theory that has only
$\Ncal=1$ supersymmetry, while the origin can be chosen to be the $\Ncal=4$ supersymmetric point (a third marginal 
direction, which preserves $\Ncal=4$ supersymmetry, corresponds to the gauge coupling $\tau$).  
 Furthermore, it has been recently shown in \cite{Doreyetal0210,Dorey03} that the S--duality of $\Ncal=4$ SYM 
does extend to an action on the \emph{vacua} of these deformed theories.

Given that these marginally deformed theories share some of the features that make the $\Ncal=4$ theory 
so special, it is natural to expect that the extension of its known string duals to these cases will
be tractable, and perhaps provide useful information.
In the AdS/CFT correspondence (see e.g. \cite{Aharonyetal00} for a review and references), $\Ncal=4$ SYM is
realised as IIB string theory on $\AdS_5\times\Srm^5$, whose isometry group ($\SU(2,2|4)$) is the same as
the superconformal group of the gauge theory. As we will briefly review later, the marginal deformations of the 
theory have also been explored in this setting, mostly in the strong coupling limit where the dual theory 
actually reduces to IIB supergravity. However, although the existence of such deformations has been established 
perturbatively, the complexity of the equations of motion of IIB supergravity has so far prevented the 
construction of an exact background dual to the marginally deformed theories\footnote{There has been recent progress
in this direction, see the Note Added at the end of our conclusions.}. 
 
The AdS/CFT correspondence is based on the equivalence of the $\Ncal=4$ theory with IIB string theory 
at \emph{strong} coupling. In \cite{Witten0312}, Witten considered instead a string theory dual of 
\emph{perturbative} $\Ncal=4$ theory, which turned out to be a topological string theory known as the B--model, 
with target space not a usual Calabi--Yau manifold but super--twistor space $\Supertwistor$. This space has
three even and four odd complex directions, which (as explained in \cite{Witten0312}) guarantees the existence
of a globally defined holomorphic volume form, which in turn is necessary for the B--model to be well defined. 

Although establishing that there exists a string theory that has the same field content as fundamental $\Ncal=4$ SYM is 
clearly remarkable, \cite{Witten0312} went further and showed that one can actually reproduce (initially
at tree level) the scattering amplitudes of the theory via a construction that is quite natural from the 
string point of view. This has led to much insight about the structure of these amplitudes and greatly 
simplified their calculation, especially when a large number of external particles is concerned 
\cite{Cachazoetal0403}\footnote{For applications and development of this formalism at tree--level, see
\cite{Zhu0403,GeorgiouKhoze0404,WuZhu0406,Benaetal0406,WuZhu0406a,Kosower0406,Georgiouetal0407,Khoze04,SuWU0409}.}.

Like $\AdS_5$ in AdS/CFT, the appearance of twistor space $\Cset\mathrm{P}^3$ here is very natural, 
as it is the space where 
the four--dimensional conformal group $\SO(2,4)\sim \SU(2,2)$ is realised in a linear way. Since this group
is unbroken for any conformal field theory, we would expect that other conformal field theories in four 
dimensions will admit a twistor string reformulation. A natural starting point, proposed in \cite{Witten0312}, 
would be to look at super--Calabi--Yau target spaces other than $\Supertwistor$, and one such case, where the target 
is a weighted projective space, was considered in \cite{PopovWolf04}. The resulting $d=4$  theories correspond 
to various self--dual truncations of $\Ncal=4$ SYM or topological $\Ncal=4$ SYM. However, it would also be 
interesting to understand how (and whether) a given theory known to be superconformal can be described using 
twistor strings.

The reason this is important lies not so much in computing scattering amplitudes in these theories (since
the approach of \cite{Cachazoetal0403} has been shown to apply to a much wider class of gauge theories than 
$\Ncal=4$ SYM) but in the insight it could provide on topological strings and their relation to gauge theory in
general. For instance, motivated by the Montonen--Olive duality of $\Ncal=4$, the authors of
\cite{NeitzkeVafa04,Nekrasovetal04} have uncovered evidence of a type of S--duality relating the 
A--model with the B--model on the same manifold (complementing the well--known mirror symmetry map between these 
models on different manifolds) along with the existence of new types of topological branes. 

As a first step in understanding how a given conformal field theory would arise in the twistor framework, we can ask 
how the abovementioned marginal deformations 
of $\Ncal=4$ SYM are encoded in the B--model picture. We will approach
this problem from the viewpoint of open/closed string theory, by considering (to first order) the effect of a particular 
closed string background field on open string correlation functions. This leads to a deformation of the action of 
holomorphic Chern--Simons theory,  which can be interpreted as the effect of introducing a \emph{non--anticommutative}
structure on some of the fermionic coordinates of $\Supertwistor$.  

The structure of our paper is the following: In section \ref{LS} we review the marginal deformations of
the $\Ncal=4$ theory, and, since we will be interested in actual perturbative calculations in these theories, 
explicitly write out their action and show how, in the same way as $\Ncal=4$ Yang--Mills, 
it can be split into a ``self--dual'' and ``non--self--dual'' part---the
first step towards reinterpretation as a twistor string theory. Section \ref{BModel} is another preparatory
section, where we review some facts about the B model on a usual, bosonic Calabi--Yau. We focus in particular
on the mixed (open/closed) amplitudes, and how they affect the target--space action of the model. After we 
discuss the generalisation of these results to the case where the Calabi--Yau is a supermanifold, in section 
\ref{TwistorStrings} we specialise to the B model on $\Supertwistor$ and identify a particular closed string
mode that induces precisely the Leigh--Strassler deformation on the self--dual part of the action. It turns out
that one can think of this deformation as introducing a very special type of non--anticommutativity on some of the
odd coordinates of $\Supertwistor$. 
 
The obvious next step is to see how the prescription of \cite{Witten0312} for the calculation of amplitudes 
in $\Ncal=4$ SYM needs to be modified to accommodate the more general case of the Leigh--Strassler theories. 
 To facilitate the reader we have inserted a section (section \ref{Analytic})
where we review the standard method and present a few illustrative examples. In section \ref{DeformedAmplitudes}
we show that the required modification of the method is simply to multiply the wavefunctions with a suitable 
star product. To check our proposal, we calculate several amplitudes to linear order in the deformation parameter 
and check that they match the ones obtained from the deformed action using Feynman diagrams. Section \ref{Comments}
 contains a preliminary discussion of higher--order terms, while in section \ref{Conclusions} we conclude 
by discussing open issues and possible extensions.

\section{Marginal Deformations of $\Ncal=4$ Super Yang--Mills} \label{LS}

Soon after it was realised that $\Ncal=4$ super Yang--Mills seemed to be a completely (UV) finite theory (see e.g. 
\cite{Sohnius85} for an account), it became
clear that it might not be the unique four dimensional theory with that property. Working in the context
of $\Ncal=1$ supersymmetry, \cite{ParkesWest84,JonesMezincescu84} obtained a set of conditions that guarantee 
finiteness at one loop for a gauge theory coupled to some matter. Considering a gauge theory with gauge group $G$ 
and matter fields $\Phi^a_I$, with $a$ labelling the representation of the gauge group and $I$ the remaining
internal indices, we can think of the cubic part of the superpotential as 
$W=\frac{1}{3!}C^{IJK}_{abc}\Phi^a_I\Phi^b_J\Phi^c_K$. Then the conditions for one--loop finiteness are
\begin{equation}\label{oneloop}
3C_2(G)=\sum_I T(R_I),\quad\text{and}\quad C^{IKL}_{acd}\overline{C}_{JKL}^{bcd}=2g^2\delta_a^{\;b}\delta^I_{\;J}T(R_I)\;.
\end{equation}
Here $C_2(G)$ is the quadratic Casimir of the group and $T(R_I)$ is the second--order Dynkin index of the representation,
defined through $\Tr(T^a_{R}T^b_{R})=T(R)\delta^{ab}$. These conditions actually suffice to show finiteness to
two loops (and vanishing of the gauge beta function to three) \cite{ParkesWest85,Grisaruetal85}, but they fail 
at higher orders for generic couplings because the three--loop anomalous dimensions are not constrained. 
However \cite{Jones86} one can imagine an iterative procedure where one chooses the dependence
of $C^{IJK}_{abc}$ on the gauge coupling at each order such that the anomalous dimensions vanish, 
which should guarantee a vanishing beta function at the next order. The resulting theory would be finite
to all orders in perturbation theory\footnote{See references in \cite{LeighStrassler95} for more on these
matters.}. 

 One obvious solution to (\ref{oneloop})  is to take the gauge group to be $\SU(N)$ (hence $C_2(G)=N$), and 
take $\Phi^a_I$ to be in
the adjoint, with $I=1,2,3$. Then the first condition is automatically satisfied. If one now chooses the interaction
coefficients to be $C^{IJK}_{abc}=g\gre^{IJK}f_{abc}$, the second condition is also satisfied. What we have constructed,
of course, is simply the $\Ncal=4$ theory. However, this choice for $C^{IJK}_{abc}$ is not the most general:
There exists a class of $\Ncal=1$ theories that satisfy the criteria of (\ref{oneloop}) and includes the 
$\Ncal=4$ theory, which we now turn to.

\subsection{The Leigh--Strassler deformation}

The first systematic treatment of marginal deformations of the $\Ncal=4$ theory appears in the work of Leigh and 
Strassler \cite{LeighStrassler95}. They realised that using symmetries and the exact $\Ncal=1$ beta functions 
given in terms of the various anomalous dimensions in the problem, one could express their vanishing in terms
of equations that were linearly dependent and thus would generically have solutions. 
Among various other examples of conformal four--dimensional theories, 
they consider a gauge theory with one $\Ncal=1$ vector superfield $V$ 
and three $\Ncal=1$ chiral superfields $\Phi_I$, with all fields in the adjoint representation of $\SU(N)$. The
form of the superpotential is 
\begin{equation} \label{WLS}
\Wcal=i \kappa \mathrm{Tr} \left[e^{i\frac{\beta}{2}}\Phi_1\Phi_2\Phi_3-e^{-i\frac{\beta}{2}}\Phi_1\Phi_3\Phi_2\right]
+\rho \mathrm{Tr}\left(\Phi_1^3+\Phi_2^3+\Phi_3^3\right)\;.
\end{equation}
In addition to the three independent couplings $\kappa,\beta,\rho$ that appear in the superpotential,
there is also the gauge coupling $\tau$. Leigh and Strassler showed that there is a 
three--complex--dimensional surface $\gamma(\tau,\kappa,\beta,\rho)=0$ in 
coupling constant space where all beta functions (and anomalous dimensions, which are all equal to $\gamma$) 
vanish and thus the corresponding theories are conformally
invariant\footnote{Note that since the vanishing of the beta functions happens at zero anomalous dimension, these
theories are indeed finite. Leigh and Strassler also consider cases where conformal invariance is restored at
non--zero anomalous dimensions for some fields. Although they are conformally invariant, those theories need not
satisfy (\ref{oneloop}).}. One of the coordinates of this space of conformal theories simply corresponds 
to the gauge coupling $\tau$, whose variation preserves $\Ncal=4$ 
supersymmetry, however turning on the other couplings breaks supersymmetry to $\Ncal=1$. The function $\gamma$ is
not known beyond one--loop (apart from the $\Ncal=4$ line $\kappa=1,\beta=\rho=0$ in suitable units), but one can argue
using (\ref{oneloop}) (see also \cite{Dorey03}) that at first order in $\beta,\rho$ the parameter $\kappa$ 
retains its $\Ncal=4$ value of $1$, so that at first order in the deformation we can add two possible chiral operators 
to the $\Ncal=4$ superpotential\footnote{We have made a slight rescaling of $\beta$ to bring the superpotential in
this form.}:
\begin{equation} \label{LSsuperpotential}
\Wcal=\Wcal_{\Ncal=4}+\beta\mathrm{Tr}\left(\Phi_1\{\Phi_2,\Phi_3\}\right)+\rho\Tr\left(\Phi_1^3+\Phi_2^3+\Phi_3^3\right)\;.
\end{equation}
It will be convenient for our purposes to consider this superpotential as a special case of a more general one, 
\begin{equation} \label{hSuperpotential}
\Wcal=\Wcal_{\Ncal=4}+\frac{1}{3!}h^{IJK}\Tr(\Phi_I\Phi_J\Phi_K)
\end{equation}
where the tensor $h^{IJK}$ is totally symmetric in its indices, and thus lies in the $\mathbf{10}$ of $\SU(3)$. 
Classically this superpotential is a marginal deformation of the $\Ncal=4$ lagrangian, for any value of
$h^{IJK}$. However asking for exact marginality requires that we make a very particular choice of $h^{IJK}$, i.e. we
need to take the nonzero components to be a linear combination of 
\begin{equation}
\begin{split} \label{hexactlymarginal}
(a)&\quad h^{123}={\beta}, \quad \text{and}\\
(b)&\quad h^{111}=h^{222}=h^{333}=\rho\;.
\end{split}
\end{equation}
Clearly these are the choices of $h^{IJK}$ that reproduce the superpotential (\ref{LSsuperpotential}), and can
also be seen to satisfy the conditions (\ref{oneloop}) with suitable normalisation of $\beta$ and $\rho$.\footnote{If 
one restricts to the case $\rho=0$, the deformation is often called the $\beta$-- or $q$--deformation in the 
literature.} Explicitly, we start by choosing $C^{IJK}_{abc}\sim h^{IJK}d_{abc}$, where $d_{abc}$ is the 
$\SU(N)$ symmetric invariant (recall $d_{acd}d^{bcd}=(N^2-4)/N\delta_a^{\;b}$) and find that the 
choices in (\ref{hexactlymarginal}) are the only ones for which $h^{IKL}\hb_{JKL}\sim \delta^I_{\;J}$. 
From the point of view of \cite{LeighStrassler95}, the reason these choices are so special is that they
preserve permutation symmetry between the chiral superfields, which is crucial in arguing that they all
have the same anomalous dimension.
In this article we will be concerned with tree--level calculations, thus for the most part we will not impose
these restrictions on $h^{IJK}$. 

A natural place to look for these marginal deformations is the AdS/CFT correspondence, where $\SU(N)$
$\Ncal=4$ SYM is realised as IIB string theory on $\AdS_5\times\Srm^5$, with $N$ units of five--form
flux through the $\Srm^5$. Since the appearance of $\AdS_5$ is crucial for the exact conformal invariance of
the dual field theory, it is expected that the $\AdS_5$ part of the geometry will be unchanged when the field
theory is deformed. Since $\Srm^5$ is rigid as an Einstein manifold, the only way the geometry can be deformed
preserving $AdS_5$ is by turning on some other matter fields in the $\Srm^5$ directions. 

In \cite{Aharonyetal02} (see also \cite{FayyazuddinMukhopadhyay02}) this problem is considered in the large $N$, 
strong coupling limit where one can work with classical
IIB supergravity on $\AdS_5\times\Srm^5$. It turns out that, apart from the already present five--form flux, 
one should also turn on (complexified) three--form flux $G_{(3)}$ in the direction of the $\Srm^5$. As expected from 
the field theory side, the IIB equations of motion can be solved \cite{Aharonyetal02}, at least to second order 
in a perturbation expansion. One expects that the deformation can be integrated to a geometry which will be an exact string
theory background, but, as mentioned earlier, the solution is not yet known\footnote{Again, see the note at the
end of the conclusions.}. A step going beyond the supergravity
limit was taken in \cite{NiarchosPrezas02}, where the $\beta$--deformation was considered in the BMN limit.

There are special points along the deformation when one can say much more. These  occur when $\rho=0$ 
while $\beta$ is a root of unity. These points \cite{BerensteinLeigh00,Berensteinetal00} have
a dual interpretation as orbifolds with discrete torsion, and furthermore noncommutativity
appears in an intriguing way as a property of the \emph{vacuum manifold} of the theory. 
The marginally deformed theories have been further studied in 
\cite{Doreyetal0210,Dorey03,Dorey04}, and several remarkable properties have been demonstrated. In particular it
was shown that (as mentioned in the introduction) the S--duality of $\Ncal=4$ extends to their space of vacua, 
and that, again for special values of $\beta$, there are also new Higgs branches on moduli space. 
These are mapped by S--duality to completely new, confining branches which appear only at the 
quantum level. 
Furthermore, at large $N$ the Higgs and confining branches can be argued to be described by Little 
String Theory \cite{Dorey04}.
Finally, the integrability properties of the deformed theories at special values of the deformation
parameter were recently explored in \cite{BerensteinCherkis04}.

\subsection{The action of the marginally deformed theories}

Since we will eventually be interested in calculating scattering amplitudes in the Leigh--Strassler deformed
theories, it is useful to write down their action in full. We start from the action
in $\Ncal=1$ superspace to connect with the discussion in the previous section, but for our purposes 
it is more convenient to work with components, so we immediately revert to component notation. The superspace
action is
\begin{equation}
S=\frac{1}{16\pi}\mathrm{Im}\left(\tau\int\diff^4 x  \diff^2 \theta \mathrm{Tr} W^{\alpha}W_{\alpha}\right)+
\int\diff^4 x \diff^2\theta \diff^2{\bar\theta} \bar{\Phi} e^{V} \Phi 
+\left(g\sqrt{2}\int\diff^4 x \diff^2 \theta \Wcal(\Phi) + h.c.\right)
\end{equation}
where $\tau$ is the complexified gauge coupling constant $\tau=\frac{\Theta_{YM}}{2\pi}+\frac{4\pi i}{g^{2}}$ and $\Wcal(\Phi)$
is as in (\ref{hSuperpotential}).  In the following we will only be interested in the perturbative aspects of the theory, 
so we drop total derivative terms by setting $\Theta_{YM}=0$.
In Euclidean space we can write the component lagrangian as (see e.g. \cite{Sohnius85}):
\begin{equation}
\begin{split}
\Lcal=&\mathrm{Tr}\left(\frac{1}{g^{2}}(\frac{1}{4}F^{2}+\tlambda\Dirac\lambda )+
(D\tphi^{I})(D\phi_{I})+\tchi^{I} \Dirac \chi_{I} 
+\sqrt{2}(\lambda [\chi_{I},\tphi^{I}] - \tlambda[\tchi^{I}, \phi_{I}])\right. \\ 
+&\frac{1}{\sqrt{2}}g \left( (\epsilon^{IJK}\chi_{I}[\phi_{J} , \chi_{K}] 
+h^{IJK}\chi_{I}\{ \phi_{J} , \chi_{K}\} ) 
-(\epsilon_{IJK} \tchi^{I}[\tphi^{J} , \tchi^{K}] 
+\hb_{IJK} \tchi^{I}\{\tphi^{J} , \tchi^{K}\})\right) \\
&\!\!\!\!\!\!\left.-\frac{1}{2}g^{2}\left([\tphi^{I} , \phi_{I}]\right)^{2}-
\frac{1}{2}g^{2} \left( \epsilon^{QJK} [ \phi_{J} , \phi_{K}]+h^{QJK} \{ \phi_{J} , \phi_{K}\} \right) 
( \epsilon _{QIL} [ \tphi^{I} , \tphi^{L}]+\hb_{QIL}\{ \tphi^{I} , \tphi^{L}\}) \right)\;.
\end{split}
\end{equation}   
In the twistor string approach to $\Ncal=4$ SYM, it is important that there is a way to split the action into
a piece which is independent of the Yang--Mills coupling constant $g$, and another which is of order $g^2$. This is
achieved by suitable rescalings of the fields. If we arrange the terms in this way, we can view the 
$g\ra0$ limit of the theory as self--dual $\Ncal=4$ SYM, which has the same field content but only a subset of the
interactions of the full theory. Then we can think of the terms of order $g^2$ as a perturbation around the self--dual
theory. The required rescalings (which treat the different helicities asymmetrically) are:
\begin{equation}
\begin{split}
&(\lambda,\tlambda)\ra (g^{\frac{1}{2}}\grl,g^{\frac{3}{2}}\tlambda),\\
&(\chi,\tchi)\ra (g^{-\frac{1}{2}}\chi,g^{\frac{1}{2}}\tchi),\\
&(\phi,\tphi)\ra (\frac{1}{\sqrt{2}}\phi,\frac{1}{\sqrt{2}}\tphi).
\end{split}
\end{equation}
As for the gauge field (see e.g. \cite{Siegel99}, p. 203) we introduce a Lagrange multiplier field 
$G^{\mu\nu}$ which is an anti--self--dual two--form. Then, up to a total derivative
term (which we drop for purposes of perturbation theory) we can replace the Yang--Mills action by the first order action: 
\begin{equation}
\int\diff^4 x \Tr\left(G_{\mn}F^{\mn}-\half g^2 G_\mn G^\mn\right)\;.
\end{equation}
After all these rescalings, the deformed lagrangian takes the form
\begin{equation} \label{LSaction}
\begin{split}
\Lcal=&\mathrm{Tr}\left[GF+ 
\tlambda \Dirac \lambda+
\frac{1}{2}(D\tphi^{I})(D\phi_{I})+
\tchi^{I}\Dirac \chi^{I}+ 
\lambda[\chi_{I},\tphi^{I}]+
\frac{1}{2}\epsilon^{IJK}\chi_{I}[\phi_{J},\chi_{K}]+
\frac{1}{2}h^{IJK}\chi_{I}\{\phi_{J},\chi_{K}\}\right.\\
-&g^{2} \left(\half G^2 + \tlambda[\tchi^{I},\phi_{I}]+
\frac{1}{2}\epsilon_{IJK}\tchi^{I}[\tphi^{J},\tchi^{K}]+
\frac{1}{2}\hb_{IJK}\tchi^{I} \{\tphi^{J},\tchi^{K}\}+\right. \\
+&\left.\left.\frac{1}{8}([\tphi^{I} , \phi_{I}])^{2}+
\frac{1}{8}(\epsilon^{QJK} [ \phi_{J} , \phi_{K}]+h^{QJK} \{ \phi_{J} , \phi_{K}\})
(\epsilon_{QIL} [ \tphi^{I} , \tphi^{L}]+\hb_{QIL}\{ \tphi^{I} , \tphi^{L}\})\right)\right]\;.
\end{split}
\end{equation}
In (\ref{LSaction}) we have collected the terms that are independent of the gauge coupling in the
first line. These define the kinetic terms and interactions of the ``self--dual'' deformed $\Ncal=4$ SYM. The
remaining terms (which include some of the Yukawa interactions and all quartic terms) are the ``non--self--dual''
terms that complete the full $\Ncal=4$ SYM action\footnote{Here we are using the notion ``self--dual'' rather loosely
to include not only the terms related to self--dual Yang--Mills by $\Ncal=4$ supersymmetry, but all terms appearing
at the same order after the rescalings above.}.

\section{The B--Model} \label{BModel}

In this section we first review some well--known facts about the B--model on a Calabi--Yau manifold 
and its target space interpretation
in terms of holomorphic Chern--Simons theory. Next we consider open/closed amplitudes in the bosonic case, and
finally the generalisation to the situation that the target space is a Calabi--Yau \emph{super}manifold. This will 
prepare us for the calculation of the effect of turning on a particular closed--string mode in the next section. 

\subsection{The B--model on a bosonic Calabi--Yau}

The topological B--model is one version of the twisted $\Ncal=2$ supersymmetric non-linear $\sigma$-model in two 
dimensions \cite{Witten88}. 
For reviews of these twisted sigma models, see \cite{Dijkgraafetal90a,Witten91,Witten9207}. 
Here we will mostly follow the notation and conventions of \cite{Hofman02}. The fields of the 
B-model are the coordinates  $\phi^{\mu}, \bar{\phi}^{\bar{\mu}}$ of the target space manifold, two twisted fermions
 $\bar{\eta}^{\bar{\mu}}, \vartheta_{\mu}$, which transform as scalars on the worldsheet, and finally a twisted 
fermion $\rho^{\mu}$, which is a one-form on the worldsheet. The action of the model is the 
following\footnote{To revert to the conventions of 
\cite{Witten91} we need to rescale $\bar{\eta}^{\bar{\mu}}\ra -\bar{\eta}^{\bar{\mu}}$, 
$\vartheta_\mu\ra -i\vartheta_\mu$, 
$\rho^{\mu}\ra i\rho^{\mu}$, and set $\frac{1}{\kappa}=t$.}: 
\begin{equation} \label{Bmodelaction}
\Scal=t\int_{\Sigma}
(g_{\mu\bar{\nu}}\diff \phi^{\mu}\ast \diff\bar{\phi}^{\bar{\nu}}-
g_{\mu\bar{\nu}}\rho^{\mu}\ast D\bar{\eta}^{\bar{\nu}})+
\frac{1}{\kappa}\int_{\Sigma}
(\rho^{\mu}D\vartheta_{\mu}
-\frac{1}{2}R^{\lambda}_{\mu\bar{\mu}\nu}\rho^{\mu}\rho^{\nu}\bar{\eta}^{\bar{\mu}}\vartheta_{\lambda})\;.
\end{equation}
Here $R_{\bar{\lambda}\lambda\bar{\mu}\mu}$ is the curvature of the target 
space K\"ahler manifold $X$, and $D$ is the target space covariant derivative. Unlike the related A--model, here
it is crucial that the target actually be a Calabi--Yau manifold, which corresponds to allowing the existence of 
a globally defined holomorphic volume form (for a Calabi--Yau three--fold this is a $(3,0)$ form).
Denoting the BRST charge by $Q$, the action is invariant under the following (on shell) BRST transformation rules:
\begin{equation} \label{BRST}
\begin{split}
[Q, \phi^\mu]&=0\\
[Q,\bar{\phi}^{\bar{\mu}}]&=\bar{\eta}^{\bar{\mu}}\\
\{Q,\bar{\eta}^{\bar{\mu}}\}&=\{Q,\vartheta_\mu\}=0 \\
\{Q,\rho^{\mu}\}&=\diff\phi^{\mu}\;.
\end{split}
\end{equation}
The B-model is a topological field theory, being independent of the complex 
structure of the Riemann surface $\Sigma$ and of the K\"ahler matric of the target space. 
In addition, it is independent of the coupling constant parameter $t$, while its dependence on $\kappa$ can 
be absorbed into a rescaling of the $\vartheta$ field. We can readily see this by rewriting the action as \cite{Hofman02}
\begin{equation}
\Scal=\frac{1}{\kappa}\int_{\Sigma}\rho^{\mu}\diff\vartheta_{\mu}+
\left\{Q, \int_{\Sigma}\left(t g_{\mu\bar{\mu}}\rho^{\mu}\ast\diff\bar{\phi}^{\bar{\mu}}-\frac{1}{2\kappa}\Gamma^{\lambda}_{\mu\nu}\rho^{\mu}\rho^{\nu}\vartheta_{\lambda}\right)\right\}\;.
\end{equation}
The fact that $t$ appears in a $Q$--exact term implies that computations can be performed 
in the large $t$ limit---the weak coupling limit of the theory---and we can expect
the results to be valid for all $t$.

The observables of the model are defined as BRST closed but not exact operators 
of the following form: 
\begin{equation} \label{observables}
V=\frac{1}{p!q!}\bar{\eta}^{\bar{\mu}_1}\cdots\bar{\eta}^{\bar{\mu}_p}{V_{\bar{\mu}_1\cdots\bar{\mu}_p}}
^{\nu_1\cdots\nu_q}\vartheta_{\nu_1}\cdots\vartheta_{\nu_q}
\end{equation}
with BRST invariance imposing $\bar{\p}V=0$ and non--exactness $V\neq\bar{\p}\Lambda$, for any $\Lambda$.   
In the large $t$ limit we can identify $\bar{\eta}^{\bar{\mu}}$ as the $(0,1)$--forms $\diff\bar{\phi}^{\bar{\mu}}$ on
$X$, and $\vartheta_\mu$ as tangent space elements $\frac{\p}{\p\phi^\mu}$. Additionally, the BRST operator 
can be identified with the $\bar{\p}$ operator on $X$. Thus we can think of 
of the observables (\ref{observables}) as elements of $\oplus_{p,q}H^{p}(X,\wedge^{q}T^{(1,0)}X)=H^{-q,p}(X) $.  

The closed string field theory of the B--model has been analysed in detail in \cite{BCOV} and was found to
describe deformations of the complex structure of the manifold. The resulting field theory was called the 
Kodaira--Spencer theory of gravity.

In the case of open strings \cite{Witten9207}, the boundary conditions 
imply that $\vartheta_\mu$ vanishes on the boundary of the worldsheet,
so the physical states are just polynomials in $\bar{\eta}^{\bar{\mu}}$. In open string field theory,  
the requirement that the ghost number equals one leaves us with just the term linear in  $\bar{\eta}^{\bar{\mu}}$:
\begin{equation}
\Acal=\bar{\eta}^{\bar{\mu}} \Acal_{\bar{\mu}}\;.
\end{equation}
Given the above--mentioned identification of $\bar{\eta}^{\bar{\mu}}$ as the $(0,1)$--forms on the target space, 
we interpret $A_{\bar{\mu}}$ as a holomorphic gauge field living on the Calabi--Yau manifold. In a straightforward manner
one can show \cite{Witten9207} that the standard cubic string field theory action reduces to holomorphic Chern-Simons:
\begin{equation} \label{HCS}
\Scal=\frac{1}{2}\int_{X}\Omega\wedge\Tr\left(\Acal\dbar\Acal+\frac{2}{3}\Acal\wedge\Acal\wedge\Acal\right)\;.
\end{equation}
Here $\Omega$ is the holomorphic $(3,0)$ form of the Calabi--Yau manifold, and the trace is over $\SU(N)$ for 
oriented strings. In modern language, $\Acal$ is thought of as the worldvolume field of a space--filling $D5$--brane
(actually a stack of such branes to account for non--abelian interactions).

\subsection{Open/Closed amplitudes} \label{OpenClosed}

As is well known, in a topological conformal field theory, we can associate to a physical state the following
three types of operators (here for the closed case):
\begin{equation}
V(P)= V ^{(0)}\;,\quad
V(C)=\int_{C}V^{(1)}\;,\quad
V(\Sigma)=\int_{\Sigma}V^{(2)}
\end{equation}
where $P$ indicates a point, $C$ a closed contour and $\Sigma$ a surface. Since $P$ is 
a point, the operators $V^{(0)}$ can be identified with the usual BRST invariant 
observables. The other operators are derived from these  through descent equations 
and are thus called descendants:\footnote{Here we already anticipated the presence of a boundary by not considering
left-- and right--movers independently.}
\begin{equation}
\begin{split}
\{Q, V^{(0)}]&=0\;,\\
\{Q, V^{(1)}]&=\diff V^{(0)}\;,\\
\{Q, V^{(2)}]&=\diff V^{(1)}\;.
\end{split}
\end{equation} 
For open strings (where the worldsheet has a boundary) we can similarly define (first) descendants of
open string states $\Acal$, the difference being that (because of cyclicity) the contour $C$ is replaced 
by an integral between two consecutive punctures on the boundary.  

These ingredients can be combined to construct open/closed correlation functions\footnote{For detailed discussions
in the context of topological strings, see \cite{HofmanMa00,Lazaroiu0010,Hofman02,Herbstetal04,KapustinRozansky04}
and references therein.}. In particular, we can define on the disk
\begin{equation}
\mathcal{F}_{p_1 p_2\cdots p_n}(t)=\ip{\Acal_{p_1}\Acal_{p_2}\mathcal{P}\int\Acal_{p_3}\cdots\int\Acal_{p_{n-1}}\Acal_{p_n}
e^{t\int_{D}V^{(2)}}}_{D}\;.
\end{equation}
Here $\mathcal{P}$ means path--ordered integration. This correlation function describes an open string $n$--point function,
deformed by the presence of a closed string background. We will now specialise to the B--model and study 
closed string deformations of the three--point open string correlation function. Recall that this is the amplitude
that specifies the cubic terms in the target space string field theory action (\ref{HCS}). 
We will also consider only the first term in the expansion of the background, so the worldsheet we will be 
interested in is the disk with three boundary punctures and one bulk puncture. This geometry has two moduli to 
be integrated over, which correspond to the position of the bulk insertion. It is known, however 
\cite{HofmanMa00,Herbstetal04}, that the resulting correlation function is equivalent (up to normalisation) to
the one with the bulk insertion fixed, and the moduli associated to the positions of two of the open string insertions.
Explicitly,
\begin{equation} \label{equivalence}
\ip{\int V^{(2)}\Acal_{p_1}\Acal_{p_2}\Acal_{p_3}}_D=\ip{V^{(0)}\Acal_{p_1}\int \Acal_{p_2}^{(1)}\int \Acal_{p_3}^{(1)}}_D\;.
\end{equation}
This expresses the deformation of the product of the open string algebra by a single closed string operator. 
Let us now calculate the second correlation function of (\ref{equivalence}) in the B--model. 
We will need the form of the first descendant boundary operators: 
\begin{equation} \label{firstdescendant}
\Acal^{(1)}= \Acal_{\bar{\mu}}(\phi, \bar{\phi})\diff \bar{\phi}^{\bar{\mu}} 
+ \p_{\mu}\Acal_{\bar{\mu}}(\phi, \bar{\phi}) \rho^{\mu} \bar{\eta}^{\bar{\mu}}\;.
\end{equation}
It is easy to see that this satisfies the relevant descent equations. 
\begin{figure}[t] 
\begin{center}
\begin{picture}(100,80)(0,15)
\Line(25,75)(50,50)\Text(23,75)[r]{$\Acal^{(1)}$} \Text(39,69)[l]{$p_3$}
\Line(75,75)(50,50)\Text(77,75)[l]{$\Acal^{(1)}$} \Text(66,59)[l]{$p_2$}
\Line(75,25)(50,50)\Text(77,25)[l]{$\Acal$} \Text(60,33)[t]{$p_1$}
\GCirc(50,50){15}{.9}
\Vertex(50,50){2}
\Photon(25,25)(50,50){-2}{5}\Text(23,25)[r]{$\phi$}
\end{picture}
\caption{A closed string contribution to the open string disk three--point amplitude.}
\end{center}
\end{figure}
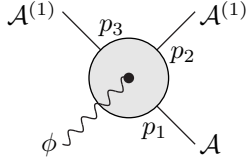
We still have to pick a particular closed string mode from the ones appearing in (\ref{observables}).
For purposes to be clear later, we choose the vertex operator 
$V^{(0)}=V \in H^{0}_{\bar{\p}}(X,\wedge^{2} T_{X})=H^{-2,0}(X)$. We follow the treatment 
of \cite{Hofman02}.
Since this vertex operator contains two $\vartheta_{\mu}$'s and they have 
no zero modes, to obtain a non-vanishing correlation function on the disk, 
one has to contract them with two $\rho^{\mu}$'s. Looking at (\ref{firstdescendant}) we see that these are provided 
by the two descendants\footnote{So we see that this closed string mode yields no corrections to the one and two point functions.}
. The correlation function is then:
\begin{equation} \label{OCcorrelator}
\begin{split}
&\langle V\Acal^{(0)}\int \Acal^{(1)}\int \Acal^{(1)}\rangle_D= \\
&\int \diff \sigma \int \diff \sigma' \langle(\frac{1}{2}V^{\mu \nu} \vartheta_{\mu} 
\vartheta_{\nu})(u)(\Acal_{\bar{\mu}} \bar{\eta}^{\bar{\mu}})(1)
(\Acal_{\bar{\nu}}\p_{\sigma}\bar{\phi}^{\bar{\nu}}+ 
\p_{\lambda}\Acal_{\bar{\nu}} \rho^{\lambda}_{\sigma}\bar{\eta}^{\bar{\nu}})(\sigma)
(\Acal_{\bar{\lambda}}\p_{\sigma'}\bar{\phi}^{\bar{\lambda}}+ 
\p_{\tau}\Acal_{\bar{\lambda}} \rho^{\tau}_{\sigma'} \bar{\eta}^{\bar{\lambda}})(\sigma') \rangle_D\;.
\end{split}
\end{equation}
We will calculate (\ref{OCcorrelator}) in the weak coupling limit $t\ra\infty$ where the path integral reduces 
to an integral over the zero modes of the fields. Here, $\sigma$ and $\sigma'$ are the coordinates of the insertions on 
the boundary (at points $p_{2}$ and $p_{3}$ in figure 1) and integration is understood to run between neighbouring 
insertions. Ordering is therefore important.
First, we integrate over the three zero modes of $\bar{\eta}^{\bar{\mu}}$. 
The result is simply $-\epsilon^{\bar{\mu}\bar{\nu}\bar{\lambda}}$ which 
provides the wedge product for the $\Acal_{\bar{\mu}}$'s.
The contraction of the $\vartheta_{\mu}$'s with the $\rho^{\lambda}_{\sigma}$'s will then give a 
term equal to $\frac{2\kappa^{2}}{2\pi(\sigma-u)2\pi(\sigma'-u)}$. This we can integrate as:
\begin{equation}
\oint \int_{1}^{\sigma} \frac{\diff\sigma\diff \sigma'}{(\sigma -u)(\sigma'-u)}=
\int_{0}^{2\pi} \int_{0}^{\varphi}  
\frac{\diff(e^{i\varphi}) \diff(e^{i\varphi'})}
{(e^{i\varphi}-u)(e^{i\varphi'}-u)}= 
\frac{1}{2}\oint \diff z \frac{1}{z-u}\oint \diff z' \frac{1}{z'-u}=-\frac{(2\pi)^2}{2}\;.
\end{equation} 
Including the overall factor of $-\frac{\kappa^2}{(2\pi)^2}$ we arrive to the conclusion that \cite{Hofman02}:
\begin{equation} \label{DeformedCubic}
\langle V \Acal^{(0)}\int \Acal^{(1)}\int\Acal^{(1)}\rangle_D=
\frac{\kappa^{2}}{2} \int_{X}\Omega \wedge \mathrm{Tr}(\Acal\wedge V^{\mu\nu}\p_{\mu}\Acal\wedge\p_{\nu}\Acal)\;.  
\end{equation}
As is obvious from the result,  the specific closed string mode provides 
a deformation of the cubic term of the holomorphic Chern-Simons action. 
More precisely, one can recognize in this term the first order correction corresponding to 
a noncommutative star product. What about the higher orders? 
In fact, as discussed in \cite{Hofman02}, higher order open/closed 
string correlators have a similar structure to those appearing in the Cattaneo--Felder model \cite{CattaneoFelder99} 
and are thus expected to give rise to the full non-commutative 
product defined by Kontsevich \cite{Kontsevich97} in the context of deformation quantisation: 
\begin{equation}\label{Kontsevich}
\begin{split} 
\Acal \ast \Acal=&\Acal\wedge\Acal+
\frac{\kappa^{2}}{2}V^{\mu\nu}\p_{\mu}\Acal\wedge\p_{\nu}\Acal+
\frac{\kappa^{4}}{8}V^{\mu\nu}V^{\lambda\tau}\p_{\mu}\p_{\lambda}\Acal\wedge\p_{\nu}\p_{\tau}\Acal+\\
&+\frac{\kappa^{4}}{12}V^{\mu\nu}(\p_{\nu}V^{\lambda\tau})
\left(\p_{\mu}\p_{\lambda}\Acal\wedge\p_{\tau}\Acal-\p_{\lambda}\Acal\wedge\p_{\mu}\p_{\tau}\Acal\right)+
\mathcal{O}(\kappa^{6}V^{3})\;.
\end{split}
\end{equation}
We should remark here that for a generic Calabi--Yau threefold the deformations we have considered are absent, 
since, by contracting with the holomorphic three--form, the dimension of $H^{-2,0}$ can be mapped 
to $h^{1,0}$ which vanishes as a consequence of the $\SU(3)$ holonomy. Thus one usually restricts 
attention to the complex structure 
deformations, which lie in $H^{-1,1}$. However, $H^{-2,0}$ can be nontrivial for special Calabi--Yau's
with even more reduced holonomy, and such cases have been studied in \cite{Kapustin0310}.

\subsection{Extension to supermanifolds}

So far we have discussed well--known facts about the B--model on an arbitrary Calabi--Yau manifold. Of course, 
for our application to $\Supertwistor$ \cite{Witten0312} we need a super--Calabi--Yau target space. In the following 
we give a brief, non--rigorous (and possibly naive) discussion of the generalisation to this case.
Since the B--model, being topological, does not
explicitly depend on the target space metric, we can (for simplicity) think of working in a metric that 
is block diagonal, with one block containing the bosonic and another containing the fermionic part\footnote{Note
that this metric does not have to be the Ricci--flat metric on the manifold. For instance, the 
super--Fubini--Study metric on $\mathbb{C}\mathrm{P}^{(3|4)}$ does not split in this way.}. Then the sigma model action will 
have a completely split form, with no mixing between the bosonic and fermionic coordinates. 

Thus, we still have the action (\ref{Bmodelaction}), where we now think of $\mu=\{i,A\}$ as a coordinate over
both the even and odd subspaces, and the various fields have either only bosonic indices $(i,\bar{i})$ or
fermionic ones $(A,\bar{A})$. The BRST rules (\ref{BRST}) are also trivially extended to this case. It follows
that the observables of the B--model on a super--Calabi--Yau now belong to 
$\oplus_{p,q,m,n}H^{p|q}(X,\wedge^{m|n}T^{(1,0)}X)$. They can be written as
\begin{equation}
V=\frac{1}{p!q!m!n!}\bar{\eta}^{\bar{i}_1}\cdots\bar{\eta}^{\bar{i}_p}\bar{\eta}^{\bar{A}_1}\cdots
\bar{\eta}^{\bar{A}_q}{V_{\bar{i}_1\cdots\bar{i}_p\bar{A}_1\cdots\bar{A}_q}}^{i_1\cdots i_m A_1\cdots A_n}
\vartheta_{i_1}\cdots\vartheta_{i_m}\vartheta_{A_1}\cdots\vartheta_{A_n}
\end{equation}
where ${V_{\bar{i}_1\cdots}}^{i_1\cdots}$ generically depends on all the supermanifold coordinates
$(Z^i,\Zb^{\bar{i}},\psi^A,\bar{\psi}^{\bar{A}})$, but has to be such that $V$ is BRST closed but not
exact. Note that the above expression is antisymmetric in the $i$ indices but \emph{symmetric} in the
$A$ indices, since $\bar{\eta}^{\bar{A}}$ and $\vartheta_{A}$ are commuting. 
Finally, the requirement that our manifold be super--Calabi--Yau boils down to the existence of a globally defined
holomorphic volume form, which, specialising to the most relevant case of $\Supertwistor$,  
 can roughly be taken to be $\Omega=\diff^3Z\diff^4\psi$. 

Turning to open strings, we can now follow \cite{Witten0312} in extending the holomorphic Chern--Simons
action (\ref{HCS}) to the specific case of $\Supertwistor$: 
\begin{equation} \label{SHCS}
\Scal=\frac{1}{2}\int_{D5}\Omega\wedge\Tr\left(\Acal\dbar\Acal+\frac{2}{3}\Acal\wedge\Acal\wedge\Acal\right)\;.
\end{equation}
It is now crucial to observe (see \cite{Witten0312} for a discussion of this) that the 
fields $\Acal(Z^i,\Zb^{\bar{i}},\psi^A)$ now live on the worldvolume 
of a $D5$--brane which is not space--filling. Rather, it is identified with the submanifold
$\bar{\psi}^{\bar{A}}=0$ within $\Supertwistor$. 
In a similar way, we can extend the discussion of open/closed amplitudes to $\Supertwistor$. We will 
say more about this case in section \ref{Corrections}, but we note here that since the volume
form is now a seven--form, the argument for triviality of deformations in $H^{-2,0}$ that 
we discussed at the end of the previous section might not go through in the same way in the
supermanifold case.

%------------------------------------------------------------------------------------------------------------

\section{Twistor String Theory} \label{TwistorStrings}

 The starting point of the twistor string programme is the observation that if the above discussion on the
B--model on a Calabi--Yau is extended to supertwistor space $\Supertwistor$, the spectrum of physical states
can be mapped (via the Penrose transform) to that of the $\Ncal=4$ theory. In the following we will review
the spectrum and open string action of the B--model on $\Supertwistor$, and then, applying the discussion in
section \ref{OpenClosed}, examine the effects of turning on a very particular closed string mode.

\subsection{The B--model on $\mathbb{C}\mathrm{P}^{3|4}$} 

Our starting point is the holomorphic Chern--Simons action (\ref{HCS}). Since we want to write things in an
 $\SU(3)\times\Urm(1)$ invariant way in preparation for breaking supersymmetry down to $\Ncal=1$, we distinguish 
the four fermionic coordinates $\psi^A$ of $\Supertwistor$ as $\psi^A=\{\xi,\psi^I\}$, where $I=1,2,3$ is the
index parametrizing the $\SU(3)$. We can now expand the superfield $\Acal(x,\psi^A)$ into components as 
\begin{equation} \label{Superfield}
\begin{split}
\Acal=&A(Z,\Zb)+\xi\lambda(Z,\Zb)+\psi^I\chi_I(Z,\Zb)+\xi\psi^I\phi_I(Z,\Zb)+\half\psi^I\psi^J\gre_{IJK}\tphi^K(Z,\Zb)\\
&+\half\xi\psi^I\psi^J\gre_{IJK}\tchi^K(Z,\Zb)+\frac{1}{3!}\psi^I\psi^J\psi^K\gre_{IJK}\tlambda(Z,\Zb)
+\frac{1}{3!}\xi\psi^I\psi^J\psi^K\gre_{IJK}G(Z,\Zb)\;.
\end{split}
\end{equation}
All fields are understood as functions of the bosonic coordinates of twistor space. As discussed in \cite{Witten0312}
these fields (which are just one--forms with certain homogeneity properties on $\Cset\mathrm{P}^3$) can be mapped
through the Penrose transform\footnote{See the appendix of \cite{Witten0312} for a review and references.}
 to fields with helicities corresponding to their homogeneity. 
According to this mapping, $A$ (having homogeneity zero) is mapped to the positive helicity part of a four--dimensional
gauge field, $\lambda,\chi_I$ give helicity $+\half$ Weyl spinors, $\phi_I,\tphi^I$ yield scalars, $\tlambda,\tchi^I$
helicity $-\half$ Weyl spinors and finally $G$ maps to the negative helicity part of the gauge field (we denote
the four--dimensional fields with the same letters as the six dimensional ones, hoping no confusion will arise).
Clearly we can group the (4d) $\{A,\lambda,\tlambda,G\}$ together to obtain an on--shell $\Ncal=1$ vector multiplet, 
and the $\{\chi_I,\phi_I,\tphi^I,\tchi^I\}$ together make up three on--shell $\Ncal=1$ chiral
multiplets (one for each value of $I$), to obtain the field content of $\Ncal=4$ SYM in $\SU(3)\times\Urm(1)$ notation. 

Expanding the holomorphic Chern--Simons action (\ref{HCS}) in components, we obtain 
\begin{equation}\label{Componentaction}
\begin{split}
\Scal_{HCS}=\int_{\mathbb{C}\mathrm{P}^3}\Omega\wedge
\Tr\left(G\wedge F+\tlambda\wedge\Dbar\grl-\tchi^I\wedge\Dbar\chi_I
+\tphi^{I}\wedge\Dbar\phi_I \right.\\
\left.-\lambda\wedge(\chi_I\wedge\tphi^I+\tphi^I\wedge\chi_I)+\gre^{IJK}\chi_I\wedge\phi_J\wedge\chi_K\right)\;.
\end{split}
\end{equation}
The first line in (\ref{Componentaction}) contains the gauge--covariantised kinetic terms, while the 
second line contains what would correspond to Yukawa--type couplings in the four dimensional theory. 
Through the Penrose transform, this action is analogous to the self--dual part of the
four--dimensional $\Ncal=4$ SYM action \cite{Witten0312}\footnote{Strictly speaking the standard Penrose transform
applies to free fields. The full, nonlinear correspondence has been analysed in \cite{PopovSamann04}.}. 
Now we would like to see how this action can be deformed. 

\subsection{Closed string corrections to holomorphic Chern--Simons} \label{Corrections}

After reviewing the correspondence between the holomorphic Chern--Simons lagrangian and the self--dual 
part of the $\Ncal=4$ lagrangian, we now turn to understanding the effect of turning on a closed--string
background as in section \ref{OpenClosed} in the context of $\Supertwistor$. 
In particular, let us consider the deformation 
(\ref{DeformedCubic}), but supersymmetrised so that all fields become superfields. The closed string modes generically
depend on all coordinates, but the open--string modes $\Acal$ living on the $D5$--brane are constrained to 
lie at $\bar{\psi}^{\bar{A}}=0$.
Also, the indices $\mu,\nu$ in 
that formula will now range over both the even and odd coordinates of $\Supertwistor$. 
To proceed, we take the indices of $V$ to both be in the fermionic directions, and in particular in the directions
$\psi^I$ corresponding to the $\SU(3)$ part. This corresponds to choosing 
\begin{equation}
V=\frac{1}{2}V^{IJ}(Z,\Zb,\psi^A,\bar{\psi}^{\bar{A}})\vartheta_I\vartheta_J
\end{equation}
as the closed string mode in (\ref{OCcorrelator}). So this is a closed string background that lives purely
in the fermionic directions of $\Supertwistor$, and explictly breaks the symmetry between $\xi$ and the other 
three $\psi^I$ components of $\psi^A$.  
Then the total target--space action, including the correction term induced by $V$, 
is\footnote{Here we have absorbed the B--model coupling $\kappa$ into the definition of $V^{IJ}$. In addition 
fermionic derivatives are defined as $\psi^I\overleftarrow{\frac{\p}{\p\psi^J}}=
\overrightarrow{\frac{\p}{\p\psi^J}}\psi^I=\delta^I_J$.}  
\begin{equation} \label{DeformedAction}
\Scal_{def}=\Scal_{HCS}+\frac{1}{2}\int_{D5}\Omega\wedge\Tr\left(V^{IJ}(Z,\Zb,\psi^A)
\Acal\wedge(\Acal\overleftarrow{\frac{\p}{\p\psi^I}})
\wedge(\overrightarrow{\frac{\p}{\p\psi^J}}\Acal)\right)\;.
\end{equation}
Note that in (\ref{DeformedAction}) $V^{IJ}$ can still depend on both the bosonic and fermionic 
coordinates $Z,\Zb,\psi^A$ of $\Supertwistor$, but not on $\bar{\psi}^{\bar{A}}$ since the brane sits
at  $\bar{\psi}^{\bar{A}}=0$. We will now make a particular choice for the 
coordinate dependence of this vertex operator, corresponding to a choice of a particular closed string background.
We thus specify that it be constant in the bosonic directions, but have a quadratic
dependence on the $\psi$ coordinates:
\begin{equation} \label{VV}
V=\frac{1}{2}\Vcal^{IJ}_{\;\;\;KL}\psi^K\psi^L\vartheta_I\vartheta_J
\end{equation}
where now $\Vcal^{IJ}_{\;\;KL}$ is a constant tensor, which has to be symmetric in its upper indices ($\vartheta_\mu$ is 
normally fermionic but its component in the odd directions $\vartheta_I$ will be bosonic) and antisymmetric in its 
lower indices. In effect, we have Taylor expanded the superfield 
$V^{IJ}(\psi^A)=V^{IJ}(\xi,\psi^I)$ in the fermionic directions, and set all coefficients to zero apart
from $\Vcal^{IJ}_{\;\;KL}$. 
There is a final step to fully specify this tensor, because it actually corresponds to
a reducible {\bf 18} of $\SU(3)$. We would like to project to an irreducible representation, so we make 
a further choice by taking:
\begin{equation} \label{Vh}
\Vcal^{IJ}_{\;\;\;KL}=h^{IJQ}\gre_{QKL}\;.
\end{equation}
Here $h^{IJQ}$ is taken to be totally symmetric, resulting in the irreducible {\bf 10} of $\SU(3)$. So we 
have finally completely specified our closed string background\footnote{It is easy to check that our final $V$
is BRST closed and non--exact.}, and we can turn to checking what, if any, 
corrections are added to the holomorphic Chern--Simons action when this mode is turned on. The additional piece 
\begin{equation} \label{hdef}
\Scal_{def}-\Scal_{HCS}=\int_{D5}\Omega\wedge\Tr\left(\frac{1}{2}h^{IJQ}\gre_{QKL}\psi^K\psi^L\Acal
\wedge(\Acal\overleftarrow{\frac{\p}{\p\psi^I}})
\wedge(\overrightarrow{\frac{\p}{\p\psi^J}}\Acal)\right)
\end{equation}
leads to several terms, most of which turn out to be zero after integration over the four fermionic 
coordinates of the $D5$--brane, even if they contain the right number of fermions.
 For instance, one of the terms that arises is\footnote{In the following two expressions we ignore the bosonic
part of $\Omega$ and focus only on the fermionic part of the measure $\diff^4\psi:=\diff\xi\diff\psi^1\diff\psi^2\diff\psi^3$.}
\begin{equation}
\begin{split}
\frac{1}{2}&\int\diff\xi\diff\psi^1\diff\psi^2\diff\psi^3 h^{IJR}\gre_{RKL}\psi^K\psi^L A\wedge (\xi\phi_I)\wedge
 (\gre_{JNQ}\psi^N\tphi^Q)\\
\;=&\frac{1}{2}\gre^{KLN} h^{IJR}\gre_{RKL} A\wedge\phi_I\wedge\tphi^Q\gre_{JNQ}
=h^{IJR} A\wedge\phi_I\wedge\tphi^Q \gre_{JRQ}
\end{split}
\end{equation}
which vanishes, exactly because $h^{IJR}$ is totally symmetric. The terms that do not vanish are
\begin{equation}
\begin{split}
\frac{1}{2}&\int\diff^{4}\psi h^{IJR}\gre_{RKL}\psi^K\psi^L\Tr(\psi^M\chi_M\wedge\chi_I\wedge(\xi\phi_J)+\psi^M\chi_M\wedge(\xi\phi_I)\wedge\chi_J
-\xi\psi^M\phi_M\wedge\chi_I\wedge\chi_J)\\
=&\frac{1}{2}\int\diff^{4}\psi h^{IJR}\gre_{RKL}\psi^K\psi^L\Tr(\psi^M\chi_M\wedge\chi_I\wedge(\xi\phi_J)+\chi_J\wedge\psi^M\chi_M\wedge(\xi\phi_I)
-\chi_I\wedge\chi_J\wedge(\xi\psi^M\phi_M))\\
=&\frac{1}{2}\int\diff^{4}\psi (\xi\psi^M\psi^K\psi^L)h^{IJR}\gre_{RKL}\Tr(-\chi_M\wedge\chi_I\wedge\phi_J+\chi_J\wedge\chi_M\wedge\phi_I
-\chi_I\wedge\chi_J\wedge\phi_R)\\
=&h^{IJM}\Tr(-\chi_M\wedge\chi_I\wedge\phi_J+\chi_J\wedge\chi_M\wedge\phi_I
-\chi_I\wedge\chi_J\wedge\phi_M)=-h^{IJK}\Tr(\chi_I\wedge\chi_J\wedge\phi_K)\\ 
=&h^{IJK}\Tr(\chi_I\wedge\phi_J\wedge\chi_K)
\end{split}
\end{equation}
where to pass from the third to the fourth line we integrated over superspace, to obtain 
$\gre^{MKL}h^{IJR}\gre_{RKL}=2h^{IJR}\delta^M_{\;R}$. So this is the only correction term 
we need to add to the B--model open string field  theory action (\ref{Componentaction}), and we are 
left with the following component action for the modes living on the bosonic part of the $D5$--brane, which is 
just $\Cset\mathrm{P}^3$:
\begin{equation}\label{Deformedcomponentaction}
\begin{split}
\Scal_{dHCS}=&\int_{\mathbb{C}\mathrm{P}^3}\Omega\wedge
\Tr\left(G\wedge F+\tlambda\wedge\Dbar\grl-\tchi^I\wedge\Dbar\chi_I
+\tphi^{I}\wedge\Dbar\phi_I \right.\\
&\qquad\qquad\qquad\;\left.-\lambda\wedge(\chi_I\wedge\tphi^I+\tphi^I\wedge\chi_I)+(\gre^{IJK}+h^{IJK})
\chi_I\wedge\phi_J\wedge\chi_K\right)\;.
\end{split}
\end{equation}
Comparing this deformed holomorphic Chern--Simons action with the four dimensional action (\ref{LSaction}) 
and repeating the arguments in \cite{Witten0312}
about the Penrose transform, we conclude that the open/closed correlation function with the particular 
choice of closed string background 
we made in (\ref{VV}) and (\ref{Vh}) ends up adding a term that is analogous to the self--dual part of 
the marginal deformation.
Observe that the new term explicitly breaks the $\SU(3)\times\Urm(1)$ symmetry, and in exactly the same way as the
deformation of the superpotential does in the four--dimensional theory. In principle one should check the 
equivalence of the two theories at the nonlinear level by repeating the analysis of \cite{PopovSamann04} in this
case.

We expect, but have not verified, that higher--point disk amplitudes (suitably regularised to account
for the contributions of the boundary of moduli space where two insertions meet) can be matched 
to amplitudes calculated from the action (\ref{Deformedcomponentaction}) using Feynman diagrams, as 
is the case for the undeformed action \cite{Witten9207}.

\subsection{Interpretation as a star product} \label{Starproductsection}

As we saw in section \ref{OpenClosed}, the bosonic deformations in $H^{-2,0}(X)$, for $X$ a Calabi--Yau,
 amount to a change in the product between open string field modes. In our case, 
following the same line of thought, we can interpret the effect of turning 
on $V^{IJ}$ (as in (\ref{DeformedAction})) as introducing non--anticommutativity between the $\psi^I$s,
leaving the product of $\psi^I$ with the bosonic coordinates and $\xi$ unaffected. 

The calculation we performed (in (\ref{OCcorrelator})) is valid only to linear order in the deformation parameter (in other
words, we have not considered disk diagrams with more than one closed string mode). For now we will consider
the implications of (\ref{hdef}) and comment on higher order corrections at the end of this section.
We will work in a non--anticommutative $\Supertwistor$, or rather on the worldvolume of
a non--space--filling $D5$ brane within $\Supertwistor$ (defined by $\bar{\psi}^I=\bar{\xi}=0$)
with coordinates satisfying the following commutation relations:  
\begin{equation}
\begin{split}
[Z,Z]=&[Z,\Zb]=[\Zb,\Zb]=0,\quad [Z,\xi]=[\Zb,\xi]=[Z,\psi^I]=[\Zb,\psi^I]=0 \\
&\{\xi,\psi^{I}\}=0,\quad\{\psi^{I},\psi^{J}\}=h^{IJQ}\epsilon_{QKL}\psi^{K}\psi^{L}\;.
\end{split}
\end{equation}
To account for the only nontrivial commutation relation, we introduce the following star product:
\begin{equation} \label{SDstarproduct}
\Acal\ast\Bcal= 
\Acal\Bcal+\Acal\overleftarrow{\frac{\p}{\p \psi^{I}}}\frac{1}{2} h^{IJQ}\epsilon_{QKL}\psi^{K}\psi^{L}
\overrightarrow {\frac{\p}{\p \psi^{J}}}\Bcal   
\end{equation}
for any superfields $\Acal(Z,\Zb,\xi,\psi^I)$ and $\Bcal(Z,\Zb,\xi,\psi^I)$, which we assume to be scalar
for simplicity. The deformed superspace thus defined is a graded Poisson structure on 
$\Supertwistor$, or put differently, a graded Lie algebra with the additional 
requirement that (here $p_\Acal$ denotes the grading of the field $\Acal$ etc.):
\begin{equation}
\begin{split}
\{\Acal \Bcal , \Ccal] =\Acal \{ \Bcal, \Ccal] + (-1)^{p_{\Bcal} p_{\Ccal}} \{\Acal, \Ccal] \Bcal\;, \\
\{\Acal, \Bcal \Ccal] = (-1)^{p_{\Bcal} p_{\Acal}} \Bcal \{ \Acal, \Ccal] + \{ \Acal, \Bcal] \Ccal\;.
\end{split}
\end{equation}  
To explicitly verify the last equations is quite straightforward. Note that in applying (\ref{SDstarproduct}), 
expressions of the form $\psi^{I}\psi^{J}\psi^{K}$ (i.e. multiplied without a $\ast$) are 
understood as being Weyl ordered (i.e. they anticommute as usual). These equations are satisfied for any choice 
of $h^{IJK}$. The validity of the Jacobi Identity however, which ensures the 
associativity of the star product, depends on the symmetry properties of $h^{IJK}$. For example, for bosonic superfields 
\begin{equation} \label{Jacobi}
[\Ccal, [\Acal, \Bcal]]+[\Acal, [\Bcal, \Ccal]] +[\Bcal, [\Ccal, \Acal]] = \mathcal{F}_{IJK} h^{IJQ} h^{KLM} 
\epsilon_{QLN} \epsilon_{MRS} \psi^{R} \psi^{S} \psi^{N} 
\end{equation}
where $\mathcal{F}_{IJK}$, which we do not write out explicitly, is a generically nonzero function that depends 
solely on various components of the superfields. We can easily deduce\footnote{To see this, note that we
can replace $\psi^R\psi^S\psi^N=\gre^{RSN}\psi^1\psi^2\psi^3$ in (\ref{Jacobi}).}
 that with $h^{IJK}$ being totally symmetric, as we have chosen it in (\ref{Vh}), the Jacobi Identity 
is automatically satisfied. 

We can now write the deformed holomorphic Chern--Simons action in terms of this star product. It is 
straightforward to verify that the action
\begin{equation}
\Scal_{def}=\half\int_{D5}\Omega\wedge\Tr\left(\Acal\ast\dbar \Acal+\frac{2}{3}\Acal\ast\Acal\ast\Acal\right)
\end{equation}
reproduces the component action (\ref{Deformedcomponentaction}). (Here we make the obvious generalisation
of (\ref{SDstarproduct}) to forms).

One might worry about the fact that in defining the star product as we did, we only took account of the terms 
linear in the deformation parameter $h^{IJK}$. We assume, as discussed in section \ref{OpenClosed} based on the 
results of \cite{Hofman02}, that the higher order open/closed correlators would yield the formula proposed by Kontsevich 
in \cite{Kontsevich97}, suitably applied to the fermionic case we consider\footnote{An extension of the 
path--integral formula of \cite{CattaneoFelder99} to the spacetime supersymmetric case can be found 
in \cite{ChepelevCiocarlie03}, however the focus there is on the non--anticommutativity properties of
the superspace coordinates, which, unlike our case,  are spacetime spinors.}. Note here that, although Kontsevich's
product was derived for the case of the deformation being a Poisson structure, it is argued in 
\cite{CornalbaSchiappa01} that the same formula would arise from disk calculations even for the non-associative case.

However it is easy to check that in our case, due to the fermionic coordinate dependence of the deformation parameter,
most of the higher order terms in Kontsevich's product immediately vanish (for instance, the second term in the 
formula (\ref{Kontsevich}) would contain four $\psi^I$'s which gives zero). As for the term containing a derivative
on $V^{IJ}$, it also vanishes by the total symmetry of $h^{IJK}$ (which, as we saw above, also guarantees the 
associativity 
of the product). We conclude that the star product (\ref{SDstarproduct}) is actually \emph{exact} and therefore encodes 
the full effect of turning on the closed string deformation (\ref{VV})-(\ref{Vh}) on the open string fields, 
and thus on the holomorphic Chern--Simons action.

Let us, in addition, remark that the deformation parameter $V^{IJ}=\Vcal^{IJ}_{\;\;\;ST}\psi^S\psi^T$, 
being a bivector, is usually associated to the inverse of the $B$ field \cite{Schomerus99,SeibergWitten9908} 
(in the regime of large $B$ where one can ignore the closed string metric). Then, the condition 
of associativity is linked to the condition that $B$ be a closed two--form. Given the 
non--invertibility of $V^{IJ}$ due to its fermionic dependence, we do not know how to relate it with 
a $B$ field in this sense.

%------------------------------------------------------------------------------------------------

\section{Analytic Amplitudes in $\Ncal=4$ Super--Yang--Mills} \label{Analytic}

In this section we will review the calculation of analytic tree--level amplitudes in the $\Ncal=4$ theory, 
as set out in \cite{Witten0312}, based on earlier work in \cite{Nair88}. We will not be very thorough since
this is well--known material (see e.g. \cite{Khoze04}). However we will do everything in an $\SU(3)\times\Urm(1)$
invariant way, since we are interested in breaking supersymmetry to $\Ncal=1$ in the next section. 

To make contact with four dimensional physics, recall that the momentum $p_{\gra\dot{\gra}}$ of a massless
particle can be decomposed as $p_{\alpha\dot{\alpha}}=\grl_\alpha\tilde{\grl}_{\dot{\alpha}}$, where 
$\grl_\alpha$ and $\tilde{\grl}_{\dot{\alpha}}$ are two--component commuting spinors called twistors. In terms
of these spinors, one can construct all fifteen generators of the four--dimensional conformal group $\SU(2,2)$. 
However, if we define the conjugate variables $\mu^{\dot{\alpha}},\mu_{\dot{\alpha}}$ through $\tilde{\grl}_{\dot{\alpha}}\ra
i\frac{\p}{\p\mu^{\dot{\alpha}}}$ and $-i\frac{\p}{\p\tilde{\grl}^{\dot{\alpha}}}\ra\mu_{\dot{\alpha}}$,
the generators of the conformal group all become first order in derivatives (see \cite{Witten0312}). The space spanned 
by the coordinates $Z=(\grl^\alpha,\mu_{\dot{\alpha}})$ and their conjugates, modulo the relation $Z\sim t Z$, is 
the twistor space we have been discussing. If the theory is \emph{super}conformal, one can also add fermionic variables
$\psi^A$, which results in supertwistor space $\Supertwistor$. In the following we use the standard notation
$\ip{12}=\gre_{\alpha\beta}\grl^{\alpha}_1\grl^{\beta}_2$ and  
$[12]=\gre^{\dot{\alpha}\dot{\beta}}\tilde{\grl}_{\dot{\alpha}1}\tilde{\grl}_{\dot{\beta}2}$.

\subsection{The prescription for analytic amplitudes} \label{Prescription}

 As observed in \cite{Witten0312}, the MHV amplitudes in 
gauge theory can be thought of as supported
on holomorphic degree one complex lines in supertwistor space $\Supertwistor$. In the topological string
approach of \cite{Witten0312}, these degree one curves are interpreted as the $D1$--branes of the B--model. 
In this context the rules given in \cite{Nair88} are derived by considering
the interaction between the fields living on the brane and background gauge field, and then considering the
contribution of a $D1$--brane background to correlation functions of these fields.
 Actually, since we will
be mostly interested in $\Ncal=4$ amplitudes containing external fields other than gluons, and these do not 
necessarily correspond to maximal helicity violation even if they are supported on degree one curves, we 
follow \cite{Khoze04} in calling all degree one amplitudes \emph{analytic}\footnote{The notation stems from the
fact that these amplitudes contain only $\ip{pq}$ contractions, not $[pq]$ ones.}. 
 So now we briefly recall the resulting prescription for the calculation of analytic amplitudes. 

First, one should consider only sub--amplitudes with a definite cyclic order, and then sum over non--cyclic orderings
at the end. To each incoming (on--shell) field we associate a wavefunction $w_i$, which is essentially the coefficient 
of that field in the expansion of the superfield $\Acal$ as given in (\ref{Superfield})\footnote{The analysis of
\cite{Witten0312} includes other factors that are useful for transforming to coordinate space, but
we will not write these out as we are only interested in momentum space amplitudes.}. For instance,
for an incoming $\tchi^I$ we take $w_i=\half\xi\psi^M\psi^N\gre_{MNI}$.

Since a particular $D1$--brane embedded in $\Supertwistor$ clearly breaks the $\SU(2,2|4)$ superconformal group, we are 
instructed to integrate over all possible such $D1$--branes. Thus we need to know their moduli space. In 
the case of degree one (which is all we will consider) the conditions for a holomorphically embedded, genus zero,
curve are 
\begin{equation}
\mu_{\dot{\alpha}}+x_{\alpha\dot{\alpha}}\grl^\alpha=0\;\; \text{and}\;\; 
\psi^A+\theta^{A}_\alpha\grl^\alpha=0\;.
\end{equation}
Since we wish to work in $\SU(3)\times\Urm(1)$ notation, we split the last equation into two:
\begin{equation} \label{psitheta}
\xi+\theta^0_\alpha\grl^\alpha=0 \;\; \text{and}\;\; \psi^I+\theta^I_\alpha\grl^\alpha=0\;.
\end{equation}
Now we can put all the above ingredients together in the formula 
\begin{equation} \label{Amplitudeformula}
\Acal_{(n)}=\int\diff^8\theta\, w_1\cdot w_2\cdots w_n \ip{J^{a_1}_1\cdots J^{a_n}_n}
\end{equation}
for an $n$--point analytic amplitude. Here the $J$'s are free--fermion currents on the $D1$--brane which 
couple to the external ($D5$--brane) fields. The path integral is over the fields living on the $D1$--brane's
worldvolume.
It is clear that the nonzero amplitudes are those that saturate the 
fermionic integral by providing precisely two $\xi$'s and six $\psi$'s. The product of the currents 
provides the gauge theory trace\footnote{To be precise, it also involves multitrace terms which we ignore.} 
and the denominator of the analytic amplitudes:
\begin{equation}
\ip{J^{a_1}(\grl_1)\cdots J^{a_n}(\grl_n)}=\Tr(T^{a_1}\cdots T^{a_n})\frac{1}{\ip{12}\cdots\ip{n1}}\;.
\end{equation}
The wavefunctions, on the other hand, give rise to the numerator of the amplitudes when integrated over
the fermionic directions. In the next subsection we 
will go through a few such analytic amplitudes, mainly in order to emphasize the differences with the deformed
theory in the next sections. Since we consider only amplitudes with specified cyclic order, we will not
write the gauge theory trace explicitly in the following.

\subsection{Some sample amplitudes} \label{UndeformedAmplitudes}

In this section we give four examples of the computation of analytic amplitudes in $\Ncal=4$ SYM, mainly in order to 
set notation and to highlight differences with the deformed amplitudes when we compute them in section 
\ref{DeformedAmplitudes}. These calculations are well--known (see the list of references in the introduction,
and \cite{Khoze04} for a review), the only significant
deviation from the literature being the $\SU(3)\times\Urm(1)$ notation. 

\paragraph{A. The amplitude $(A_1A_2G_3G_4)$} \mbox{}

This is the standard MHV four point amplitude that appears in pure Yang--Mills theory. According to the superfield
expansion (\ref{Superfield}) the positive helicity gluon $A$ has no $\psi$ dependence, while the negative helicity
gluon $G$ is found at order $(\xi\psi^3)$, so the prescription outlined above dictates that we write
\begin{equation}
\Acal_{(4)}^{AAGG}=\int\diff^8\theta 
\frac{1}{3!3!}(\xi_3\psi^I_3\psi^J_3\psi^K_3)(\xi_4\psi^L_4\psi^M_4\psi^N_4)\gre_{IJK}\gre_{LMN}
\frac{1}{\ip{12}\ip{23}\ip{34}\ip{41}}\;.
\end{equation}
To integrate over the fermions, we should convert the $\xi,\psi$ coordinates to the $\theta$ ones. To do this we 
use equation (\ref{psitheta}). The $\xi$'s clearly give a unique choice:
\begin{equation}
\xi_3\xi_4=\theta^0_\alpha\grl_3^\alpha\theta^0_\beta\grl_4^\beta=(\theta^0)^2\gre_{\alpha\beta}\grl_3^\alpha\grl_4^\beta
=(\theta^0)^2\ip{34}\;.
\end{equation}
So integration gives $\int\diff^2\theta^0\,\xi_3\xi_4=\ip{34}$.
As for the $\psi$'s, we have various ways to contract the indices before integrating, and this will lead
to more structure. For later use, let us do the integration in steps. One way to do the contraction 
of the $\SU(3)$ indices is the following:
\begin{equation}
\Wicktall{15mm}\psi^I_3\Wickunder{15mm}\psi^J_3\Wick{15mm}\psi^K_3\psi^L_4\psi^M_4\psi^N_4
=\psi^1_3\psi^2_3\psi^3_3\psi^1_4\psi^2_4\psi^3_4
(\delta^{IL}\delta^{JM}\delta^{KN})\;.
\end{equation}
Integrating this over the six remaining $\theta$ coordinates gives $(-\delta^{IL}\delta^{JM}\delta^{KN})\ip{34}^3$
There are five more contractions to do, which give the result\footnote{Here and in the following we use 
 $\int\diff^8\theta=\int\diff^2\theta^0\diff^6\theta$ to denote integration over all fermionic moduli, and define  
$\int\diff^6\theta=\diff^2\theta^1\diff^2\theta^2\diff^2\theta^3$ for just the $\theta^I$'s.} 
\begin{equation} \label{AAGGcontractions}
\int\diff^6\theta \psi^I_3\psi^J_3\psi^K_3\psi^L_4\psi^M_4\psi^N_4=-\gre^{IJK}\gre^{LMN}\ip{34}^3\;.
\end{equation}
Putting all the factors together, we obtain (with an extra minus from anticommuting $\xi_4$ to the left)
\begin{equation} \label{AAGG}
\Acal_{(4)}^{AAGG}=\frac{\ip{34}^4}{\ip{12}\ip{23}\ip{34}\ip{41}}
\end{equation}
which is the familiar formula for this MHV amplitude.

\paragraph{B. The amplitude $(\tchi^I_1\tchi^J_2\tphi^K_3)$} \mbox{}

This is the analytic amplitude that is conjugate to the $\chi\chi\phi$ vertex of the self--dual theory. The relevant 
part of the amplitude is
\begin{equation} 
\Acal_{(3)}^{\tchi\tchi\tphi}=\int\diff^8\theta \frac{1}{2^3}(\xi_1\psi^M_1\psi^N_1)(\xi_2\psi^P_2\psi^Q_2)
(\psi^R_3\psi^S_3)\gre_{MNI}\gre_{PQJ}\gre_{RSK}\;.
\end{equation}
As before, the $\xi$ integration gives a factor of $\ip{12}$, but this time the integration over the $\psi$'s gives
\begin{equation}
\int\diff^6\theta\psi^M_1\psi^N_1\psi^P_2\psi^Q_2\psi^R_3\psi^S_3=
-\gre_{XYZ}\gre^{XMN}\gre^{YPQ}\gre^{ZRS}\ip{12}\ip{23}\ip{31}\;.
\end{equation}
Thus, after integrating over all odd coordinates, we obtain
\begin{equation}
\Acal_{(3)}^{\tchi\tchi\tphi}=
-\frac{1}{8}\gre_{XYZ}\gre^{XMN}\gre^{YPQ}\gre^{ZRS}\gre_{MNI}\gre_{PQJ}\gre_{RSK}\frac{\ip{12}^2\ip{23}\ip{31}}{\ip{12}\ip{23}\ip{31}}
= -\gre_{IJK}\ip{12}\;.
\end{equation}
Looking at (\ref{LSaction}), we find agreement with the corresponding term in the lagrangian (note the
overall minus sign in (\ref{LSaction})).

\paragraph{C. The amplitude $(\tlambda_1,\tchi_2^I,\phi_{J,3})$} \mbox{}

This is the amplitude conjugate to the $\grl\chi\tphi$ vertex in the self--dual action. Its structure
is 
\begin{equation}
\Acal_{(3)}^{\tlambda\tchi\phi}=
\int\diff^8\theta\frac{1}{3!2}(\psi^P_1\psi^Q_1\psi^R_1)(\xi_2\psi^M_2\psi^N_2)(\xi_3\psi^J_3)\gre_{PQR}\gre_{MNI}
\frac{1}{\ip{12}\ip{23}\ip{31}}\;.
\end{equation}
Integrating over the $\psi$ coordinates results in
\begin{equation}
\int\diff^6\theta \psi^P_1\psi^Q_1\psi^R_1\psi^M_2\psi^N_2\psi^J_3=
\gre^{PQR}\gre^{MNJ}\ip{12}^2\ip{13}
\end{equation}
which, inserted in the amplitude, gives
\begin{equation}
\Acal_{(3)}^{\tlambda\tchi\phi}=\frac{1}{3!2}\gre^{PRQ}\gre^{MNJ}\gre_{PQR}\gre_{MNI}
\frac{\ip{12}^2\ip{13}\ip{23}}{\ip{12}\ip{23}\ip{31}}
=-\delta_I^{\;\;J}\ip{12}\;.
\end{equation}
Note that the spinor product that appears is the same as in the previous example, and also the normalisation of
the amplitude is exactly what we would expect from the corresponding term in the lagrangian (\ref{LSaction}). The
fact that these amplitudes are equal (apart from their $\SU(3)\times\Urm(1)$ structure) is of course a
consequence of $\Ncal=4$ supersymmetry.

\paragraph{D. The amplitude $(\chi_{I,1}\chi_{J,2}\tchi^K_3\tchi^L_4)$} \mbox{}

As  a final, more interesting example, let us calculate this particular four point amplitude. It goes like
\begin{equation} \label{chifouramp}
\Acal_{(4)}^{\chi\chi\tchi\tchi}=\int\diff^8\theta \frac{1}{2\cdot2}\psi^I_1\psi^J_2(\xi_3\psi^M_3\psi^N_3)(\xi_4\psi^P_4\psi^Q_4)\gre_{MNK}\gre_{PQL}\frac{1}{\ip{12}\ip{23}\ip{34}\ip{41}}\;.
\end{equation}
Here we have an important difference from the previous examples we considered, in that there
are three possible contractions of the momenta when integrating over the $\theta$ coordinates. 
They are 
\begin{equation} \label{chifourabc}
\begin{split}
(a)&\;\int\diff^6\theta \Wick{4mm}\psi^I_1\psi^J_2\Wick{10mm}\psi^M_3\Wickunder{10mm}\psi^N_3\psi^P_4\psi^Q_4 
+\{ ^{M\leftrightarrow N}_{\!\,\, P\leftrightarrow Q} \}=
\delta^{IJ}(\delta^{MQ}\delta^{NP}-\delta^{MP}\delta^{NQ})\ip{12}\ip{34}^2\\ \\
(b)&\; \int\diff^6\theta \Wick{24mm}\psi^I_1\Wickunder{4mm}\psi^J_2\psi^M_3\Wickunder{4.5mm}\psi^N_3\psi^P_4\psi^Q_4 
+\{ ^{M\leftrightarrow N}_{\!\,\, P\leftrightarrow Q} \}
=\left(\delta^{IJ}(\delta^{MP}\delta^{NQ}-\delta^{MQ}\delta^{NP})-\gre^{IMN}\gre^{JPQ}\right)\ip{23}\ip{34}\ip{41}\\ \\
(c)& \; \int\diff^6\theta \Wick{9mm}\psi^I_1\Wickunder{14.5mm}\psi^J_2\psi^M_3\Wick{10mm}\psi^N_3\psi^P_4\psi^Q_4 
+\{ ^{M\leftrightarrow N}_{\!\,\, P\leftrightarrow Q} \}
=\left(\delta^{IJ}(\delta^{MP}\delta^{NQ}-\delta^{MQ}\delta^{NP})-\gre^{IPQ}\gre^{JMN}\right)\ip{13}\ip{24}\ip{34}\;.
\end{split}
\end{equation}
On each line we have summed over the different permutations that are related by  
$M\leftrightarrow N$ and $P\leftrightarrow Q$. In fact, the three momentum structures in (\ref{chifourabc}) 
are not independent, but are related through the Schouten identity 
(see e.g. \cite{Siegel99}, p. 141) 
\begin{equation} \label{Schouten}
\langle pq\rangle\langle rs \rangle+\langle qr\rangle\langle ps \rangle+\langle rp\rangle\langle qs \rangle=0\;.
\end{equation}
Applying this to the third factor in (\ref{chifourabc}), in the form $\ip{13}\ip{24}=\ip{12}\ip{34}-\ip{23}\ip{41}$, 
we get just two factors, 
\begin{equation} \label{twofactors}
\int\diff^6\theta\psi^I_1\psi^J_2\psi^M_3\psi^N_3\psi^P_4\psi^Q_4=
-\gre^{IPQ}\gre^{JMN}\ip{12}\ip{34}^2+(\gre^{IPQ}\gre^{JMN}-\gre^{IMN}\gre^{JPQ})\ip{23}\ip{34}\ip{41}\;.
\end{equation}
Now we can insert this result in (\ref{chifouramp}), and soon obtain
\begin{equation} \label{chifourfinal}
\Acal_{(4)}^{\chi\chi\tchi\tchi}=-\delta^I_{\;\;L}\delta^{J}_{\;\;K}\frac{\ip{34}^2}{\ip{23}\ip{41}}
-\left[\gre^{IJQ}\gre_{QKL}\right]\frac{\ip{34}}{\ip{12}}\;.
\end{equation}
Since this is just a four--point amplitude, it is easy to calculate it using standard field theory Feynman 
diagrams. We find that there are two contributions, as shown in figure 2.
\begin{figure}[ht] \label{Diagrams}
\begin{center}
\begin{picture}(300,50)(0,0)
\put(-30,0){
\Vertex(50,25){2} \Vertex(100,25){2}
\SetColor{BrickRed}
\DashLine(50,25)(100,25){1}
\SetColor{Blue}
\Line(25,10)(50,25)
\Line(25,40)(50,25)
\Line(100,25)(125,40)
\Line(100,25)(125,10)
\Text(20,40)[r]{$\chi_{I,1}$}\Text(20,10)[r]{$\chi_{J,2}$}
\Text(130,10)[l]{$\tchi^K_3$}\Text(130,40)[l]{$\tchi^L_4$}
\Text(60,27)[b]{$\phi_Q$}\Text(90,27)[b]{$\tphi^Q$}
\Text(75,1)[c]{\bf{(a)}}
}
\put(170,0){
\Vertex(50,25){2} \Vertex(100,25){2}
\SetColor{BrickRed}
\Photon(50,25)(100,25){2}{5}
\SetColor{Blue}
\Line(25,10)(50,25)
\Line(25,40)(50,25)
\Line(100,25)(125,40)
\Line(100,25)(125,10)
\Text(20,40)[r]{$\tchi^L_4$}\Text(20,10)[r]{$\chi_{I,1}$}
\Text(130,10)[l]{$\chi_{J,2}$}\Text(130,40)[l]{$\tchi^K_3$}
\Text(60,29)[b]{$A$}\Text(90,29)[b]{$A$}
\Text(75,1)[c]{\bf{(b)}}
}
\end{picture}
\caption{The two Feynman diagrams that contribute to tree--level $(\chi\chi\tchi\tchi)$ scattering.}
\end{center}
\end{figure}
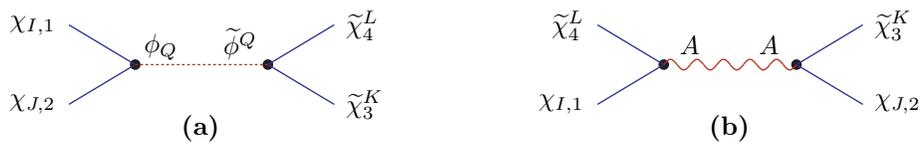
Again, we can verify the answer (\ref{chifourfinal}) by explicitly computing these Feynman diagrams. It
will be important later to be able to distinguish the contributions of these two diagrams through their
difference in  momentum structure, since only one of them (the one with scalar exchange) is expected to change 
when we deform the superpotential as in (\ref{hSuperpotential}).

%-----------------------------------------------------------------------------------------

\section{Calculation of Analytic Amplitudes in the Deformed Theory} \label{DeformedAmplitudes}

In section \ref{Starproductsection} we found that the deformation from the self--dual part of the
$\Ncal=4$ action to that of the marginally deformed theory can be described by a simple star product, which
encodes the non--anticommutativity of three of the fermionic coordinates of $\Supertwistor$. Motivated
by the appearance of non--anticommutativity, in this section we will give a heuristic prescription for
modifying the calculation of tree--level analytic amplitudes (as presented in the previous section) in order to 
directly compute amplitudes in the marginally deformed theories. We then check our proposal via several examples.

\subsection{Extension of the Star Product} \label{Extension}

Our prescription should be rather obvious: The $\psi$'s appearing in the calculation of amplitudes should
now be thought of as non--anticommuting, and thus should be multiplied with a suitable star product. The first obvious
choice is to multiply them using the same star product (\ref{SDstarproduct}). However some thought quickly shows
that (\ref{SDstarproduct}) cannot be the whole story: As can be inferred from the action (\ref{LSaction}),
generic amplitudes in the non--self dual deformed theory will involve not only the tensor $h^{IJK}$ but also its complex
conjugate $\hb_{IJK}$, and the star product (\ref{SDstarproduct}) would not be able to produce such terms. We propose
the following generalisation of (\ref{SDstarproduct}):
\begin{equation} \label{Starproduct}
f(\psi_1)\ast g(\psi_2)=f(\psi_1)g(\psi_2)+\half \Vcal^{IJ}_{\;\;\;KL}
\left(f(\psi_1)\overleftarrow{\frac{\p}{\p\psi^I_1}}\right)\psi^K_1
\psi^L_2\left(\overrightarrow{\frac{\p}{\p\psi^J_2}}g(\psi_2)\right)
\end{equation}
where $\psi_1$ and $\psi_2$ are the fermionic coordinates of two different wavefunctions, and 
the tensor $\Vcal^{IJ}_{\;\;\;KL}$ is defined to be
\begin{equation} \label{htensor}
\Vcal^{IJ}_{\;\;\;KL}=\left(h^{IJQ}\gre_{QKL}+\gre^{IJQ} \hb_{QKL}\right)\;.
\end{equation}
When acting between functions of the same coordinate $\psi$, this reduces to the star product (\ref{SDstarproduct}) 
we obtained in holomorphic Chern--Simons theory. A few simple calculations using this star product are summarized 
in the appendix. 

Clearly we have not derived (\ref{Starproduct}), but introduced it based on the fact that (as we will show)
it works. To understand 
how topological strings might lead to such a product, we would certainly need a better understanding of the behaviour 
of a $D1$--brane in the particular closed string background  we turned on in section \ref{Corrections}. 
It is important to note that the star product, as defined in (\ref{Starproduct}), is not complete, 
since it could well include terms of higher order in $\Vcal$. If we ignore
these possible terms, the star product is not associative 
(e.g. $(\psi^I_1\ast\psi^J_2)\ast\psi^K_3\neq\psi^I_1\ast(\psi^J_2\ast\psi^K_3)$), yet the non--associativity only
shows up in the terms of second order and higher in $\Vcal$. Although in this article we are only interested in the terms
linear in $\Vcal$, we will make some preliminary comments on the higher order terms in the next section. 

The implications of making the odd coordinates of superspace non--anticommutative (and the way this non--anticommutativity
arises from string theory considerations) have been much explored 
recently in the literature \cite{Klemmetal01,deBoeretal03,OoguriVafa0303,OoguriVafa0303,Seiberg03,Ferraraetal03}. 
In fact, in \cite{SaemannWolf04} this idea
is explored in the context of $\Ncal=4$ SYM, necessarily using the equations of motion due to the lack of a 
covariant superspace formulation of $\Ncal=4$ supersymmetry. The type of non--anticommutativity we have found differs
in some crucial aspects from those previously discussed. Clearly, here it is not the four--dimensional superspace that
becomes non--anticommutative, but the fermionic directions of $\Supertwistor$. Also, 
in our case the non--anticommutativity
parameter turns out to be coordinate dependent (though it depends only on the odd coordinates). We do not
know whether it is possible to map this non--anticommutativity to four dimensional superspace, though clearly
given the relation $\psi^I=\theta^I_\alpha\grl^\alpha$ for degree one curves, we could perhaps think of attributing
the non--anticommutativity to the $\theta$'s rather than the $\psi$'s. Even if this makes sense, however, it does not
seem to generalise to non--analytic amplitudes, since in that case the relation between $\theta^I_\alpha$ 
and $\psi^I$ is not linear.

To proceed, let us associate, as in section \ref{Prescription}, to each component in the superfield 
expansion (\ref{Superfield}) a wavefunction $w_i$, which we again take to be the coefficient in front of the component. 
In terms of these wavefunctions, 
the prescription we propose for calculating analytic amplitudes in the marginally deformed theory is
\begin{equation} \label{DeformedFormula}
\Acal_{(n)}=\int\diff^8\theta\, w_1\ast w_2\ast\cdots\ast w_n \frac{1}{\ip{12}\ip{23}\cdots\ip{n1}}\;.
\end{equation}
As mentioned, the only difference from the standard formula (\ref{Amplitudeformula})
is that we now multiply the wavefunctions with the star product (\ref{Starproduct}).
Note that since the twistor space superfields $\Acal$ are taken to be Weyl--ordered, the same holds for the 
corresponding wavefunctions $w_i$. This means that we need not consider star 
products between $\psi$'s belonging to the same wavefunction\footnote{Another way of seeing this is that the $\psi^I$s 
in the same wavefunction are at the same point, so the second part of (\ref{htensor}) acting on them gives zero,
but they also  always come antisymmetrised with an $\gre$--symbol, so the first part is also trivial.}.

It remains to check that our star product leads to the correct amplitudes for the Leigh--Strassler theories.
Since there are amplitudes with different numbers of star products to compute, in the following we start 
from the simplest case and move up to more complicated ones. To actually compare with field theory Feynman
diagrams  we restrict to low--point amplitudes, but, since the positive helicity gluons $A$ do not depend
on the $\psi^I$'s, and so do not affect the star products we calculate, we can trivially add any number 
of them to our amplitudes and obtain the full MHV $n$--point series.

\subsection{Products of the form $(\psi\psi\psi)\ast(\psi\psi\psi)$} \label{SecA}

This product would contribute to the purely gluonic MHV amplitude $\Acal_{(4)}^{AAGG}$, but also to
various other amplitudes 
like $\Acal_{(3)}^{A\grl \tlambda G}$ and $\Acal_{(4)}^{\grl\grl\tlambda\tlambda}$. Clearly these amplitudes, which come purely 
from the gauge part of the action, do not feel the deformation of the superpotential and should be the same 
as in the $\Ncal=4$ theory. To make the formulas shorter, we introduce the notation
\begin{equation}
\Psi_A^{IJKMNP}:=\psi^I_3\psi^J_3\psi^K_3\psi^M_4\psi^N_4\psi^P_4\;.
\end{equation}
It is easy to find the result of integrating this over the $\theta^I$ coordinates of superspace (see
(\ref{AAGGcontractions})):
\begin{equation}
\int\diff^6\theta\,\Psi_A^{IJKMNP}=-\gre^{IJK}\gre^{MNP}\ip{34}^3\;.
\end{equation}
In this notation, calculation of the star product gives 
\begin{equation}
(\psi^I_3\psi^J_3\psi^K_3)\ast(\psi^M_4\psi^N_4\psi^P_4)\gre_{IJK}\gre_{MNP}=\Psi_A^{IJKMNP}\gre_{IJK}\gre_{MNP}
+\frac{9}{2} \Vcal^{KM}_{\;\;\;XY}\Psi_A^{IJXYNP}\gre_{IJK}\gre_{MNP}\;.
\end{equation}
Let us now use this result in the calculation of the amplitude $\Acal_{(4)}^{AAGG}$. 
\begin{equation}
\begin{split}
\frac{1}{3!3!}\int \diff^8\theta (\xi_3\psi_3^I\psi_3^J\psi_3^K)
\ast(\xi_4\psi^M_4\psi^N_4\psi^P_4)\gre_{IJK}\gre_{MNP}\frac{1}{\ip{12}\ip{23}\ip{34}\ip{41}}\\
=-\int\diff^6\theta\left[\Psi_A^{IJKMNP}
+\frac{9}{2} \Vcal^{KM}_{\;\;\;XY}\Psi_A^{IJXYNP}\right]\gre_{IJK}\gre_{MNP}
\frac{\ip{34}}{\ip{12}\ip{23}\ip{34}\ip{41}}\;.
\end{split}
\end{equation}
Doing the superspace integration  gives 
\begin{equation}
\Acal_{(4)}^{AAGG}=\frac{1}{3!3!}\left[\gre^{IJK}\gre^{MNP}+\frac{9}{2}\Vcal^{KM}_{\;\;\;XY}\gre^{IJX}\gre^{YNP}\right]
\gre_{IJK}\gre_{MNP}\frac{\ip{34}^4}{\ip{12}\ip{23}\ip{34}\ip{41}}\;.
\end{equation}
The term linear in $\Vcal$ vanishes, since we have $\Vcal^{KM}_{\;\;\;XY}\delta^X_K\delta^Y_M=\Vcal^{KM}_{\;\;\;KM}=0$, 
thus the only contribution comes from the first term. So we have reproduced the standard gluonic MHV 
result (\ref{AAGG}). 

Changing the positions of the $\xi$'s gives amplitudes related to this one by $\Ncal=1$ supersymmetry, and can 
be seen to receive no $\Vcal$--corrections either. 
So we conclude that, as expected from the field theory, amplitudes containing the structure 
$(\psi\psi\psi)\ast(\psi\psi\psi)$ are identical to the $\Ncal=4$ ones.

\subsection{Products of the form $\psi\ast(\psi\psi)\ast(\psi\psi\psi)$} \label{SecB}

Depending on where the $\xi$'s are placed, this structure contributes to three--point amplitudes
like $(\phi\tphi G)$, $(\phi\tchi\tlambda)$ and $(\chi\tchi G)$, and also to four--point amplitudes
like $(\grl\phi\tphi\tlambda)$. As these amplitudes may get contributions only from D--terms, we do not
expect them to change when we turn on $\Vcal$. We again define the notation
\begin{equation} \label{PsiB}
\Psi_B^{IJKMNP}:=\psi^I_1\psi^J_2\psi^K_2\psi^M_3\psi^N_3\psi^P_3\;,\qquad 
\int\diff^6\theta\,\Psi_B^{IJKMNP}=\gre^{IJK}\gre^{MNP}\ip{13}\ip{23}^2\;.
\end{equation}
Using this notation, we can now calculate
\begin{equation}
\begin{split}
\psi^I_1&\ast(\psi^J_2\psi^K_2)\ast(\psi^M_3\psi^N_3\psi^P_3)\gre_{JKL}\gre_{MNP}=
\Psi_B^{IJKMNP}\gre_{JKL}\gre_{MNP}+\\ \\
&+\frac{1}{2}\left(6\Vcal^{KM}_{\;\;\;XY}\Psi_B^{IJXYNP}+
3\Vcal^{IM}_{\;\;\;XY}\Psi_B^{XJKYNP}+
2\Vcal^{IJ}_{\;\;\;XY}\Psi_B^{XYKMNP}\right)\gre_{JKL}\gre_{MNP}   \\ \\
&+3\Vcal^{IJ}_{\;\;\;XY}\left(\Vcal^{KM}_{\;\;\;ST}\Psi_B^{XYSTNP}
+\Vcal^{YM}_{\;\;\;ST}\Psi_B^{XSKTNP}
+\Vcal^{XM}_{\;\;\;ST}\Psi_B^{SYKTNP}\right)\gre_{JKL}\gre_{MNP}\;.
\end{split}
\end{equation}
From (\ref{PsiB}) we see that integration over $\psi$ will essentially introduce a factor of $\gre^{IJK}\gre^{MNP}$. 
Then we can easily show that both the terms linear and the terms quadratic in $\Vcal$ vanish in this case
also. For instance, take the term
\begin{equation}
3\Vcal^{KM}_{\;\;\;XY}\Psi_B^{IJXYNP}\gre_{JKL}\gre_{MNP}\ra3 \Vcal^{KM}_{\;\;\;XY}\gre^{IJX}\gre^{YNP}\gre_{JKL}\gre_{MNP}
=6 \Vcal^{KM}_{\;\;\;XM}\gre^{IJX}\gre_{JKL}=0\;.
\end{equation}
Again, this is just as it is expected to be from the field theory side.

\subsection{Products of the form $(\psi\psi)\ast(\psi\psi)\ast(\psi\psi)$} \label{SecC}

This case is more interesting. This structure appears in the three--point amplitude $(\tchi\tchi\tphi)$, which,
as can be seen from the form of the Leigh--Strassler action (\ref{LSaction}), \emph{does} change in the 
deformed theory. Thus we expect the action of the star product to be nontrivial for this amplitude. As before, we
define
\begin{equation}
\Psi_C^{MNPQUV}:=\psi^M_1\psi^N_1\psi^P_2\psi^Q_2\psi^U_3\psi^V_3, \quad 
\int\diff^6\theta\,\Psi_C^{MNPQUV}=-\gre_{XYZ}\gre^{XMN}\gre^{YPQ}\gre^{ZUV}\ip{12}\ip{23}\ip{31}
\end{equation}
and find the star product
\begin{equation}
\begin{split}
&(\psi^M_1\psi^N_1)\ast(\psi^P_2\psi^Q_2)\ast(\psi^U_3\psi^V_3)\gre_{MNI}\gre_{PQJ}\gre_{UVK}=
\left(\Psi_C^{MNPQUV}+\right. \\ \\
&\left.+2\left[\Vcal^{NP}_{\;\;\;ST}\Psi_C^{MSTQUV}+\Vcal^{QU}_{\;\;\;ST}\Psi_C^{MNPSTV}
+\Vcal^{NU}_{\;\;\;ST}\Psi_C^{MSPQTV}\right]\right) \gre_{MNI}\gre_{PQJ}\gre_{UVK}+\mathcal{O}(\Vcal^2)\;.
\end{split}
\end{equation}
The first line contains the term with no $\Vcal$'s, which contributes to the $\Ncal=4$ $(\tchi\tchi\tphi)$ amplitude, 
as we saw in section \ref{UndeformedAmplitudes}. We can now calculate the same amplitude in the deformed theory:
\begin{equation}
\begin{split}
\Acal_{(3)}^{\tchi\tchi\tphi}=&\int\diff^8\theta\frac{1}{8} (\xi_1\psi^M_1\psi^N_1)\ast(\xi_2\psi^P_2\psi^Q_2)
\ast(\psi^R_3\psi^S_3)\gre_{MNI}\gre_{PQJ}\gre_{RSK}\frac{1}{\ip{12}\ip{23}\ip{31}}=\\
&=-(\gre_{IJK}+\hb_{IJK})\ip{12}
\end{split}
\end{equation}
just as expected from the deformed action (\ref{LSaction}).

\subsection{Products of the form $\psi\ast\psi\ast\psi\ast(\psi\psi\psi)$} \label{SecD}

This product contributes to amplitudes like $(\chi\chi\phi G)$ and $(\chi\phi\phi\tlambda)$, which contain 
vertices that are affected by the marginal deformation. Defining the notation
\begin{equation}
\Psi_D^{IJKMNP}:=\psi^I_1\psi^J_2\psi^K_3\psi^M_4\psi^N_4\psi^P_4,\quad 
\int\diff^6\theta\,\Psi_D^{IJKMNP}=\gre^{IJK}\gre^{MNP}\ip{24}\ip{34}\ip{41}\;,
\end{equation}
up to linear order in the deformation we have
\begin{equation}
\begin{split}
\psi^I_1\ast\psi^J_2&\ast\psi^K_3\ast(\psi^M_4\psi^N_4\psi^P_4)\gre_{MNP}=\Psi_D^{IJKMNP}\gre_{MNP}+\\ \\
+&\frac{3}{2}\left(\Vcal^{KM}_{\;\;\;ST}\Psi_D^{IJSTNP}+\Vcal^{JM}_{\;\;\;ST}\Psi_D^{ISKTNP}
+\Vcal^{IM}_{\;\;\;ST}\Psi_D^{SJKTNP}\right)\gre_{MNP}\\ \\
+&\half\left(\Vcal^{JK}_{\;\;\;ST}\Psi_D^{ISTMNP}+\Vcal^{IK}_{\;\;\;ST}\Psi_D^{SJTMNP}
+\Vcal^{IJ}_{\;\;\;ST}\Psi_D^{STKMNP}\right)\gre_{MNP}\;.
\end{split}
\end{equation}
It is easy to see that integration over the $\theta$'s will be such that the terms in the
second line will always give factors like $\Vcal^{KM}_{\;\;\;SM}$ and thus vanish. So the only contribution 
comes from the terms in the third line, where they combine to give 
\begin{equation}
\half 6\left(\Vcal^{JK}_{\;\;\;ST}\gre^{IST}+\Vcal^{IK}_{\;\;\;ST}\gre^{SJT}+\Vcal^{IJ}_{\;\;\;ST}\gre^{STK}\right)
=6h^{IJK}\;.
\end{equation}
Taking as our example the amplitude $(\chi_I\phi_J\phi_K\tlambda)$, we find
\begin{equation}
\begin{split}
\Acal_{(4)}^{\chi\phi\phi\tlambda}=&
\int\diff^8\theta \frac{1}{3!}\left(\psi^I_1\ast(\xi_2\psi^J_2)\ast(\xi_3\psi^K_3)
\ast(\psi^M_4\psi^N_4\psi^P_4)\gre_{MNP}\right)\frac{1}{\ip{12}\ip{23}\ip{34}\ip{41}}=\\
&=-\left(\gre^{IJK}+h^{IJK}\right)\frac{\ip{24}}{\ip{12}}\;.
\end{split}
\end{equation}
This is indeed the result we expected, as can be seen from the corresponding gauge theory Feynman diagram shown in figure 3.
\begin{figure}[ht] \label{chiphiphitlambdaDiagram}
\begin{center}
\begin{picture}(120,50)(0,0)
\put(0,0){
\Vertex(50,25){2} \Vertex(80,25){2}
\SetColor{BrickRed}
\Line(50,25)(80,25)
\SetColor{Blue}
\DashLine(35,10)(50,25){1}
\Line(35,40)(50,25)
\Line(80,25)(95,40)
\DashLine(80,25)(95,10){1}
\Text(30,40)[r]{$\chi_{I,1}$}\Text(30,10)[r]{$\phi_{J,2}$}
\Text(100,10)[l]{$\phi_{K,3}$}\Text(100,40)[l]{$\tlambda_4$}
\Text(53,28)[bl]{$\chi$}\Text(77,28)[br]{$\tchi$}
}
\end{picture}
\caption{ $(\chi\phi\phi\tlambda)$ scattering.}
\end{center}
\end{figure}
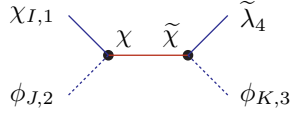

\subsection{Products of the form $\psi\ast\psi\ast(\psi\psi)\ast(\psi\psi)$} \label{SecE}

 As mentioned briefly at the end of section \ref{UndeformedAmplitudes}, this case should be more interesting
because the amplitudes it contributes to (like $(\chi\chi\tchi\tchi)$ and $(\phi\phi\tphi\tphi)$) are made 
up of more than one Feynman diagram
on the gauge theory side, and not all of these diagrams should change in the marginal theory. We take as an
example the amplitude $(\chi\chi\tchi\tchi)$. It is given by
\begin{equation}
\Acal_{(4)}^{\chi\chi\tchi\tchi}=\int\diff^8\theta\frac{1}{4}
\psi^I_1\ast \psi^J_2\ast(\xi_3\psi^M_3\psi^N_3)\ast(\xi_4\psi^P_4\psi^Q_4)\gre_{MNK}\gre_{PQL}
\frac{1}{\ip{12}\ip{23}\ip{34}\ip{41}}\;.
\end{equation}
Let us again define $\Psi_E^{IJKLMN}:=\psi^I_1\psi^J_2\psi^K_3\psi^L_3\psi^M_4\psi^N_4$. Calculating the star 
product to linear order, we find
\begin{equation}
\begin{split}
\psi^I_1\ast&\psi^J_2\ast(\psi^M_3\psi^N_3)\ast(\psi^P_4\psi^Q_4)\gre_{MNK}\gre_{PQL}=\left(\Psi_E^{IJMNPQ}
+\half\Vcal^{IJ}_{\;\;\;ST}\Psi_E^{STMNPQ}+\Vcal^{JM}_{\;\;\;ST}\Psi_E^{ISTNPQ}+\right.\\ \\
&\left.+\Vcal^{IM}_{\;\;\;ST}\Psi_E^{SJTNPQ}+\Vcal^{JP}_{\;\;\;ST}\Psi_E^{ISMNTQ}+\Vcal^{IP}_{\;\;\;ST}\Psi_E^{SJMNTQ}
+2\Vcal^{NP}_{\;\;\;ST}\Psi_E^{IJMSTQ}\right)\gre_{MNK}\gre_{PQL}\;.
\end{split}
\end{equation}
Now consider integrating this over superspace. As discussed in some detail in section \ref{UndeformedAmplitudes}.D,
this integration naturally splits up into two parts, one per underlying Feynman diagram (see (\ref{twofactors})). 
So if we want to compute corrections to the gluon exchange diagram (figure 2.a), we simply need to substitute
each occurence of $\Psi_E$ above by the corresponding product of antisymmetric tensors. Explicit calculation shows
that in this particular case all contributions linear in $\Vcal$ cancel, as expected because this diagram contains 
no vertices arising from the superpotential. 
 As for the scalar exchange diagram (figure 2.b), substituting the analogous expression for each $\Psi_E$, we 
eventually find that the terms linear in $\Vcal$ sum up to $-4\Vcal^{IJ}_{\;\;\;KL}$. 
So the overall amplitude for this example becomes
\begin{equation} \label{chifourdeformed}
\Acal_{(4)}^{\chi\chi\tchi\tchi}=-\delta^I_{\;\;L}\delta^{J}_{\;\;K}\frac{\ip{34}^2}{\ip{23}\ip{41}}
-\left[\gre^{IJQ}\gre_{QKL}+h^{IJQ}\gre_{QKL}+\gre^{IJQ}\hb_{QKL}\right]\frac{\ip{34}}{\ip{12}}
\end{equation}
exactly as we would expect from the field theory side. 

It is worth remarking that, although as we just saw the product $\psi\ast\psi\ast(\psi\psi)\ast(\psi\psi)$ does
result in terms of order $\Vcal$, the very similar--looking product $\psi\ast(\psi\psi)\ast\psi\ast(\psi\psi)$ does
not give such terms. This is just as well, because it contributes (for instance) to the 
amplitude $(\phi\tphi\phi\tphi)$, which in the gauge theory arises either from gluon exchange or from 
the D--term quartic vertex in the lagrangian (\ref{LSaction}), and so will not be affected by changes to the
superpotential.

\subsection{Higher--point functions} \label{SecF}

The two products that we have not analysed yet are $\psi\ast\psi\ast\psi\ast\psi\ast(\psi\psi)$ and
$\psi\ast\psi\ast\psi\ast\psi\ast\psi\ast\psi$. They clearly contribute to analytic amplitudes that are five--point
or higher. Computation of the corrections to these amplitudes is straightforward but tedious, so, as a final check
on our method,  we will simply do an example and check that it matches field theory expectations to linear order in
the deformation parameter. 

The example we take is the five--point amplitude $\Acal_{(5)}(\chi_I\chi_J\phi_K\phi_L\tphi^P)$. This corresponds to
\begin{equation} \label{Afive}
\Acal_{(5)}^{\chi\chi\phi\phi\tphi}=\int\diff^8\theta \half\left(\psi^I_1\ast\psi^J_2\ast(\xi_3\psi^K_3)\ast(\xi_4\psi^L_4)
\ast(\psi^M_5\psi^N_5)\gre_{MNP}\right)\frac{1}{\ip{12}\cdots\ip{51}}\;.
\end{equation}
The first task is to identify the various momentum structures that arise. Applying all possible contractions
on the six $\psi$'s, we get six different spinor products, however by (sometimes repeated) use of the Schouten
identity (\ref{Schouten}) they can be reshuffled into only three, resulting in 
\begin{equation}
\begin{split}
\int\diff^6\theta \psi^I_1\psi^J_2\psi^K_3\psi^L_4(\psi^M_5\psi^N_5)\gre_{MNP}=
&-\gre^{IMN}\gre^{JKL}\ip{12}\ip{35}\ip{45}+\gre^{IJK}\gre^{MNL}\ip{25}\ip{34}\ip{51}\\
&+(\gre^{IJK}\gre^{MNL}-\gre^{IJL}\gre^{MNK})\ip{23}\ip{45}\ip{51}\;.
\end{split}
\end{equation}
Substituting this result in (\ref{Afive}), and recalling the factor of $\ip{34}$ from the $\xi$ integration, 
we obtain three momentum structures, which can be seen to correspond to the following three Feynman diagrams:
\begin{figure}[ht] \label{FiveptDiagrams}
\begin{center}
\begin{picture}(400,90)(0,0)
\put(0,0){
\SetColor{Blue}
\Line(25,35)(40,50) \Text(23,33)[r]{$\chi_{I,1}$}
\DashLine(25,65)(40,50){1}\Text(23,67)[r]{$\tphi^P_5$}
\DashLine(70,50)(70,75){1} \Text(72,77)[b]{$\phi_{L,4}$}
\DashLine(100,50)(115,65){1}\Text(117,67)[l]{$\phi_{K,3}$}
\Line(100,50)(115,35) \Text(117,33)[l]{$\chi_{J,2}$}
\SetColor{BrickRed}
\Line(40,50)(100,50) \Text(43,53)[bl]{$\lambda$}\Text(67,53)[br]{$\tlambda$}
\Text(73,53)[bl]{$\tchi$}\Text(97,53)[br]{$\chi$}
\SetColor{Black}
\Vertex(40,50){2} \Vertex(70,50){2} \Vertex(100,50){2}
\Text(70,10)[c]{\bf{(a)}}
}
\put(150,0){
\SetColor{Blue}
\DashLine(25,35)(40,50){1} \Text(23,33)[r]{$\phi_{K,3}$}
\Line(25,65)(40,50)\Text(23,67)[r]{$\chi_{J,2}$}
\Line(70,50)(70,75)\Text(72,77)[b]{$\chi_{I,1}$}
\DashLine(100,50)(115,65){1}\Text(117,67)[l]{$\tphi^P_5$}
\DashLine(100,50)(115,35){1}\Text(117,33)[l]{$\phi_{L,4}$}
\SetColor{BrickRed}
\Line(40,50)(70,50)\Photon(70,50)(100,50){1}{4}
\Text(43,53)[bl]{$\chi$}\Text(67,53)[br]{$\tchi$}
\Text(73,53)[bl]{$A$}\Text(97,53)[br]{$A$}
\SetColor{Black}
\Vertex(40,50){2} \Vertex(70,50){2} \Vertex(100,50){2}
\Text(70,10)[c]{\bf{(b)}}
}
\put(300,0){
\SetColor{Blue}
\Line(25,35)(40,50) \Text(23,33)[r]{$\chi_{J,2}$}
\Line(25,65)(40,50) \Text(23,67)[r]{$\chi_{I,1}$}
\SetColor{BrickRed}
\DashLine(40,50)(70,50){1}
\Text(43,53)[bl]{$\phi$}\Text(67,53)[br]{$\tphi$}
\SetColor{Blue}
\DashLine(70,50)(70,70){1} \Text(72,72)[b]{$\tphi^P_5$}
\DashLine(70,50)(90,50){1} \Text(92,52)[l]{$\phi_{L,4}$}
\DashLine(70,50)(70,30){1} \Text(72,27)[t]{$\phi_{K,3}$}
\SetColor{Black}
\Vertex(40,50){2} \Vertex(70,50){2}
\Text(60,10)[c]{\bf{(c)}}
}
\end{picture}
\caption{The three amplitudes that contribute to tree--level $(\chi\chi\phi\phi\tphi)$ scattering.}
\end{center}
\end{figure}
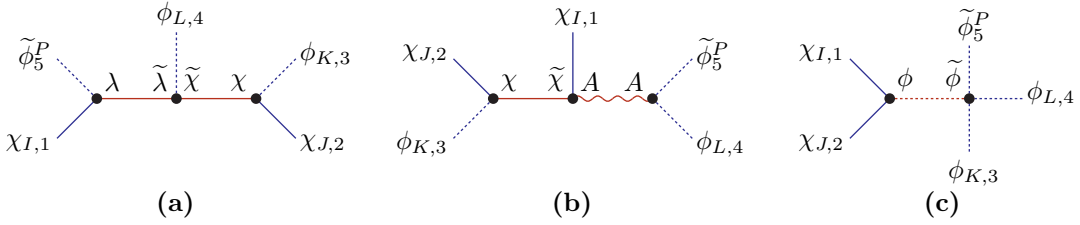

Let us now concentrate on the last of these diagrams, which involves both a three-point and a four--point
vertex. Introducing
\begin{equation}
\Psi_F^{IJKLMN}:=\psi_1^I\psi_2^J\psi_3^K\psi_4^L\psi_5^M\psi_5^N\;,
\end{equation}
computing the star product in (\ref{Afive}) and keeping just the linear terms, we have
\begin{equation}
\begin{split}
\psi^I_1&\ast\psi^J_2\ast\psi^K_3\ast\psi^L_4\ast(\psi^M_5\psi^N_5)\gre_{MNP}=
\Psi_F^{IJKLMN}\gre_{MNP}+
\frac{1}{2}\left(2\Vcal^{LM}_{\;\;\;ST}\Psi_F^{IJKSTN}+2\Vcal^{KM}_{\;\;\;ST}\Psi_F^{IJSLTN}\right.\\ \\
&+2\Vcal^{JM}_{\;\;\;ST}\Psi_F^{ISKLTN}+2\Vcal^{IM}_{\;\;\;ST}\Psi_F^{SJKLTN}
+\Vcal^{KL}_{\;\;\;ST}\Psi_F^{IJSTMN}+\Vcal^{JL}_{\;\;\;ST}\Psi_F^{ISKTMN}+\\ \\
&\left. +\Vcal^{IL}_{\;\;\;ST}\Psi_F^{SJKTMN}
+\Vcal^{JK}_{\;\;\;ST}\Psi_F^{ISTLMN}+\Vcal^{IK}_{\;\;\;ST}\Psi_F^{SJTLMN}+\Vcal^{IJ}_{\;\;\;ST}\Psi_F^{STKLMN}\right)\gre_{MNP}
+\mathcal{O}(\Vcal^2)\;.
\end{split}
\end{equation}
For the spinor product we want, integration over the $\theta$'s 
converts $\Psi_F^{IJKLMN}\ra(\gre^{IJK}\gre^{LMN}-\gre^{IJL}\gre^{KMN})$. The final
result is
\begin{equation}
\int\diff^6\theta\psi^I_1\ast\psi^J_2\ast\psi^K_3\ast\psi^L_4\ast(\psi^M_5\psi^N_5)\gre_{MNP}
=\left(2 \Vcal^{LK}_{\;\;\;SP}\gre^{IJS}+2(h^{IJK}\delta^{L}_{\;\;P}-h^{IJL}\delta^{K}_{\;\;P})\right)\ip{23}\ip{45}\ip{51}
\end{equation}
Writing this result in a slightly more suggestive way, and substituting in (\ref{Afive}), we conclude that 
\begin{equation}
\Acal_{(5c)}^{\chi\chi\phi\phi\tphi}=\left(h^{IJQ}\gre^{LKR}\gre_{QRP}+\gre^{IJQ}(h^{RLK}\gre_{RPQ}+\gre^{RLK}\hb_{RPQ})\right)
\frac{1}{\ip{12}}\;.
\end{equation}
These are exactly the terms we expect to find from the field theory point of view at linear order in the
deformation.

\section{Comments on higher order terms} \label{Comments}

In the previous section we saw, through numerous examples, that our prescription for calculating analytic
amplitudes correctly reproduces the field theory results in the marginally deformed $\Ncal=4$ theory, to
linear order in the deformation parameter $\Vcal$. How about higher orders in $\Vcal$?
It is straightforward to check that for the amplitudes in sections \ref{SecA}--\ref{SecD}, the 
higher order terms are either absent or vanish upon integration over superspace, consistent with 
field theory expectations. However the
amplitude considered in section \ref{SecE} should have contributions of second order (specifically, a
term proportional to $h^{IJQ}\hb_{QKL}$). Calculation shows that apart from the expected terms, 
we obtain some extra ones, of the form $\gre^{IJQ}\gre_{KLP}h^{MNQ}\hb_{MNP}$.

A mismatch at second order is perhaps to be expected, given that in our calculations, we have not included 
possible higher order terms in the star product (\ref{Starproduct}). An obvious guess for the second order terms
would be to consider an extension of Kontsevich's formula (\ref{Kontsevich}), as in \cite{ChepelevCiocarlie03}. 
As we saw in section 
\ref{Starproductsection}, adding such terms to the star product in the self--dual part of the theory gave no
contribution. Here, we have products between $\psi$'s of different particles, so these terms could
in principle be there. In any case, since for analytic amplitudes we never need to multiply more than six $\psi$'s,
and thus there is only a finite amount of terms that need be fixed in the star product, the issue is not really
whether we can find a proper prescription, but to understand how it would arise from the underlying theory.
Such an understanding would certainly be useful in extending our method to non--analytic amplitudes. 

A related issue concerns \emph{exactly} marginal deformations. All our calculations have assumed a general 
form for the parameter $h^{IJK}$, rather than the specific choice in (\ref{hexactlymarginal}). If we would
like to restrict ourselves to the exactly marginal case, apart from making this choice we know that at second
order in $\beta$ and $\rho$ we would have to add other operators to the superpotential, corresponding to changing
the coefficient of the $\gre^{IJK}$ term. Also, in a generic basis, the gauge coupling $g$ will depend on all
the other couplings in the theory as well. Still, by convenient rescalings of the fields, these effects will only
appear in the non--self--dual part of the action. Assuming the mapping of the self--dual piece of the
deformed action to the deformed holomorphic Chern--Simons theory, this implies that at higher orders we need 
not consider more general corrections to the latter theory, but just modifications to the prescription
for calculating amplitudes (which probes the non--self--dual part of the gauge theory)\footnote{We would
expect that our method would correspond to the theory in which, after suitable rescalings, the
coefficient of the non--self--dual $\gre_{IJK}$ coupling ($|\kappa\cos\frac{\beta}{2}|^2$ in the notation
of (\ref{WLS})) is set to one.}.

It is worth remarking that the unexpected terms at higher orders are such that if we choose $h^{IJK}$ to
be of exactly marginal form, they reduce to expressions proportional to $\gre^{IJQ}\gre_{QKL}$, which could
be thought of as arising from the $\gre^{IJK}$ term in the superpotential. However, one would expect that
since we are working at tree level, no conditions need to be imposed on the coefficient of that term for 
the gauge theory calculations to agree with the twistor ones. In other words, it would be strange to see 
any effects of exact marginality in our calculation. Could tree--level (in the sense of genus zero $D1$--instantons) 
twistor string theory know about one--loop gauge theory results? A more thorough analysis of the higher--order terms
should resolve this question.

\section{Conclusions} \label{Conclusions}

In this article we identified a closed--string deformation to the target space open--string action of the B--model
on supertwistor space that corresponds to adding a marginal deformation to the corresponding four--dimensional
field theory. The undeformed case is simply $\Ncal=4$ Super--Yang--Mills \cite{Witten0312}, while the
deformed theory belongs to the class of $\Ncal=1$ conformal field theories studied by 
Leigh and Strassler \cite{LeighStrassler95}. 

 We saw how this deformation affects the self--dual part of the theory, by adding appropriate terms to
the action of holomorphic Chern--Simons theory. These terms can be understood in a geometric way as a 
consequence of non--anticommutativity in some of the fermionic directions of $\Supertwistor$. 
As for the non--self--dual part, we found how the prescription of \cite{Nair88,Witten0312} for calculating analytic 
amplitudes in $\Ncal=4$ SYM needs to be modified to apply to the deformed four--dimensional gauge
theories. In a similar way to the self--dual case, the only modification is that we have to introduce an appropriate star
product between the wavefunctions that enter the formulas of \cite{Nair88,Witten0312}. 

However, we have shown that our prescription reproduces the expected gauge theory amplitudes only at linear
order in the deformation parameter. As discussed in the previous section, we expect that a slight 
modification of our method will be adequate to correctly account for the quadratic and higher terms.  

There are various possibilities for future work. A first task would be to understand how to compute
non--analytic tree--level amplitudes in this theory. In the original approach of \cite{Witten0312} (see also
\cite{Roibanetal0402,RoibanVolovich0402,Roibanetal0403}), these amplitudes are supported on holomorphic curves of
degree two and higher embedded in $\Supertwistor$, and thus integration over their moduli space involves
more than eight $\theta$ coordinates. Is non--anticommutativity of the $\psi$'s still all that is needed to obtain 
these amplitudes? One would also  expect that the equivalence \cite{Gukovetal0404}  between this 
connected prescription and the disconnected one of \cite{Cachazoetal0403} would persist in the marginally deformed
theories, but perhaps one should check whether the proof goes through in the same way for our non--anticommutative 
$\Supertwistor$. 
  
More interesting is the extension of our results to loop amplitudes. As we have discussed, at the classical level
any superpotential of the form (\ref{LSsuperpotential}), with $h^{IJK}$ an arbitrary symmetric tensor, will
preserve conformal invariance. In accordance with this fact, we find no obstruction to deforming the $\Ncal=4$
theory with this superpotential in our tree--level calculations. 
However, quantum--mechanically things are very different: The tensor $h^{IJK}$ should be constrained to be of the
form (\ref{hexactlymarginal}), for the deformation to be \emph{exactly} marginal. Any other value of $h^{IJK}$ would give a 
non--conformal four--dimensional theory and would thus presumably not be describable in terms of a string thery on 
$\Supertwistor$. Unfortunately, at the moment the prescription for calculation of loop diagrams from twistor space 
\cite{Cachazoetal0406,Brandhuberetal0407,Cachazoetal0409,LuoWen0410,Benaetal0410,Cachazo0410}
has not been derived in a completely fundamental way from string theory, and perhaps extending our results to loops
will have to wait for a better understanding of this issue. 

It is understood \cite{Witten0312,Giombietal0405,BerkovitsWitten04} that the \emph{closed} B--model on 
$\Supertwistor$ contains information about conformal $\Ncal=4$ supergravity in four dimensions\footnote{This 
correspondence has also been extended to other super--Calabi--Yau manifolds \cite{Ahn0409}.}. So the
closed--string mode that we have employed to bring about the marginal deformation in the gauge theory may
correspond to a particular field in conformal $\Ncal=4$ supergravity (and thus the deformation could perhaps 
be thought of as spontaneous $\Ncal=4\ra\Ncal=1$ supersymmetry breaking in $\Ncal=4$ SYM coupled to $\Ncal=4$
conformal supergravity). However, the matching of the B--model closed--string states with the fields of 
conformal $\Ncal=4$ supergravity in \cite{BerkovitsWitten04} does not seem (at first sight) to include such a 
mode\footnote{In particular, one would expect \cite{Aharonyetal00} this mode to belong to 
the {\bf 45} of $\SU(4)$ which contains the {\bf 10} of $\SU(3)$.}, and it would
be important to establish whether it is indeed part of the four dimensional supergravity theory or not. 

A related issue concerns the open twistor string formulation of \cite{Berkovits0402,BerkovitsMotl04}, which is
believed to be equivalent to the B--model approach that we have been discussing. Recall that, despite being
an open string theory, that model also encodes information about conformal supergravity in four dimensions. The vertex 
operators that correspond to supergravity in that theory have been matched with some of those of the closed B--model 
in \cite{BerkovitsWitten04}. Since the particular mode (\ref{VV}) does not seem to appear in this mapping, we
do not yet know how the deformation would arise in the approach of \cite{Berkovits0402} (would perhaps considering
a closed string completion of that formalism provide this state?)\footnote{Similarly one could 
look for this mode within the alternative formulation of \cite{Siegel0404}.}. In any case, one could think of how 
to deform the open string field theory star product of \cite{BerkovitsMotl04} to account for non--anticommuting
odd coordinates.

It will be interesting to understand how these deformations manifest themselves in other proposed 
topological string models for $\Ncal=4$ SYM \cite{NeitzkeVafa04,Nekrasovetal04,AganagicVafa04,KumarPolicastro04,
SinkovicsVerlinde04}, like the A--model
on $\Supertwistor$ or its mirror B--model on super--ambitwistor space (the quadric in 
$\Cset \mathrm{P}^{3|3}\times\Cset \mathrm{P}^{3|3}$). 

In this article we have concentrated on the Leigh--Strassler theories in their superconformal phase, 
where no fields have expectation values. As discussed earlier, these theories also have a very interesting vacuum structure,
including (apart from their Coulomb branch) Higgs and even confining phases \cite{Doreyetal0210,Dorey03,Dorey04}
for particular values of the deformation parameters. 
Is there an extension of the twistor string formalism to account for those cases also? This question has
been raised already in \cite{Witten0312} about (the presumably simpler case of) the Coulomb branch of $\Ncal=4$
SYM, but to our knowledge has not been addressed yet. It is worth remarking, however, that in the case 
of the Higgs and confining phases of the $\beta$--deformed theories, Dorey \cite{Dorey03,Dorey04} has recently argued that
(in certain limits) they can be described by a Little String Theory. Perhaps the appearance of another 
six--dimensional string theory here is not altogether coincidental.    

In summary, we have described the marginal deformations of $\Ncal=4$ SYM in the context of the twistor B--model. 
It turned out that the deformation corresponds not to a change in the geometry of $\Supertwistor$, but
is a consequence of making it non--anticommutative in a certain way. There are several unexplored issues which 
we hope to address in future work.

\paragraph{Note Added}
Since the first version of this paper, the understanding of the marginally deformed theories
at strong coupling (and large N) has greatly increased thanks to the work of Lunin and 
Maldacena \cite{LuninMaldacena05}, who constructed the supergravity dual of the $\beta$--deformation. 
Although our work is valid in the opposite regime, we believe that it will
be very useful to explore further the connections between these two string theoretical approaches to 
the marginal deformations of $\Ncal=4$ SYM.

\paragraph{Acknowledgments} 
We wish to thank B. Carlini--Vallilo, S. Giombi, R. Ionas, R. Ricci, D. Robles--Llana,
M. Ro\v{c}ek, W. Siegel and D. Trancanelli for useful discussions and comments. We would also like to 
thank the participants 
of the second Simons Workshop in Mathematics and Physics at Stony Brook for creating a 
stimulating environment
from which we benefited greatly. M. K. acknowledges support through NSF award PHY--0354776.

\appendix

\section{Examples of star product calculations} 

Since the star product defined in section \ref{Extension} is slightly unfamiliar because of the $\psi$--dependence of
the deformation parameter, in this appendix we do a few simple examples. First, note that as
defined in (\ref{htensor}), the tensor $\Vcal^{IJ}_{\;\;\;KL}$ satisfies 
$\Vcal^{IJ}_{\;\;\;KL}=-\Vcal^{JI}_{\;\;\;LK}$ and also $\Vcal^{IM}_{\;\;\;JM}=0$. The basic star product
between two $\psi$'s is
\begin{equation} 
\psi^I_1\ast\psi^J_2=\psi^I_1\psi^J_2+\half \Vcal^{IJ}_{\;\;\; KL}\psi^K_1\psi^L_2
\end{equation}
It is certainly worth remarking (and perhaps relevant for discussions of integrability) 
that similar quadratic products appear in the context of quantum groups, see
\cite{deBoeretal03} for a short discussion in the context of non--anticommutative superspace. 
A more complicated example is the following:
\begin{equation}
\begin{split}
\psi^I_1\ast\left(\psi^K_2\psi^L_2\right)&\gre_{KLM}=
\psi^I_1\psi^K_2\psi^L_2\gre_{KLM}+\half\left(\Vcal^{IK}_{\;\;\; RS}\psi^R_1\psi^S_2\psi^L_2
-\Vcal^{IL}_{\;\;\; RS}\psi^R_1\psi^S_2\psi^K_2\right)\gre_{KLM}\\
&=\psi^I_1\psi^K_2\psi^L_2\gre_{KLM}+\half\left(\Vcal^{IK}_{\;\;\; RS}\psi^R_1\psi^S_2\psi^L_2
+\Vcal^{IL}_{\;\;\; RS}\psi^R_1\psi^K_2\psi^S_2\right)\gre_{KLM}\\
&=\psi^I_1\psi^K_2\psi^L_2\gre_{KLM}+\Vcal^{IK}_{\;\;\; RS}\psi^R_1\psi^S_2\psi^L_2\gre_{KLM}\
\end{split}
\end{equation}
To go from the first line to the second one we simply anticommuted $\psi^S_2$ and $\psi^K_2$, which we are
allowed to do since all expressions are now Weyl ordered. In passing from the second to the third line
we made use of the antisymmetric tensor $\gre_{KLM}$ to interchange the $K,L$ indices of the  second term in
the parenthesis. This brings in a minus sign which is cancelled when we anticommute the $\psi$'s, so we 
see that this term is exactly equal to the first term in the parenthesis. 

A final useful formula that can be derived from the definition (\ref{Starproduct}) is
\begin{equation}
(\psi^I_1\psi^J_1)\ast(\psi^M_2\psi^N_2)\gre_{IJK}\gre_{MNP}=
\psi^I_1\psi^J_1\psi^M_2\psi^N_2\gre_{IJK}\gre_{MNP}+2 \Vcal^{JM}_{\;\;\;RS}\psi^I_1\psi^R_1\psi^S_2\psi^N_2\gre_{IJK}\gre_{MNP}
\end{equation}
As discussed in the main text, the definition (\ref{Starproduct}) leads to a non--associative star product. We
can see this by calculating: 
\begin{equation} \label{nonassociativity}
(\psi^I_1\ast\psi^J_2)\ast\psi^K_3-\psi^I_1\ast(\psi^J_2\ast\psi^K_3)=\frac{1}{4}
\left(\Vcal_{MP}^{IJ}\Vcal_{NQ}^{MK}+\Vcal_{NM}^{IJ}\Vcal_{PQ}^{MK}
-\Vcal_{PM}^{JK}\Vcal_{NQ}^{IM}-\Vcal_{MQ}^{JK}\Vcal_{NP}^{IM}\right)\psi^N_1\psi_2^P\psi_3^Q
\end{equation}
Thus non--associativity arises only at second order in $\Vcal$.

\bibliography{TwistorRefs}
\bibliographystyle{JHEP}

\end{document}